\documentclass[traditabstract,twocolumn]{aa}

\usepackage{txfonts}
\usepackage{graphicx}	
\usepackage{amsmath}	
\usepackage{amssymb}	
\usepackage{gensymb}
\usepackage{xcolor}     
\usepackage{soul}       
\usepackage{array}
\usepackage{xspace}
\usepackage[colorlinks=true,linkcolor=blue,citecolor=blue, urlcolor=blue]{hyperref}
\bibpunct{(}{)}{;}{a}{}{,}

\usepackage{fontawesome}  

\newcommand{\glee}{\textsc{GLEE}\xspace}
\newcommand{\lenstronomy}{\textsc{Lenstronomy}\xspace}
\newcommand{\vkl}{\textsc{VKL}\xspace}
\newcommand{\herculens}{\textsc{Herculens}\xspace}
\newcommand{\qlens}{\textsc{QLens}\xspace}
\newcommand{\molet}{\textsc{MOLET}\xspace}
\newcommand{\coolest}{\textsc{COOLEST}\xspace}
\newcommand{\nifty}{\textsc{NIFTy}\xspace}

\begin{document}


\title{Exploiting the diversity of modeling methods\\to probe systematic biases in strong lensing analyses}

\titlerunning{Modeling methods and lensing degeneracies}

\author{A.~Galan\inst{\ref{tum},\ref{mpa},\ref{epfl}}, G.~Vernardos\inst{\ref{lehman},\ref{amnh}}, Q.~Minor\inst{\ref{bmcc},\ref{amnh}}, D.~Sluse\inst{\ref{uliege}}, L.~Van de Vyvere\inst{\ref{uliege}} \and M.~Gomer\inst{\ref{uliege}}}

\institute{
Technical University of Munich, TUM School of Natural Sciences, Department of Physics, James-Franck-Str 1, 85748 Garching,
Germany \label{tum}
\goodbreak
\and
Max-Planck-Institut für Astrophysik, Karl-Schwarzschild-Str. 1, 85748 Garching, Germany \label{mpa}
\goodbreak
\and
Institute of Physics, Laboratory of Astrophysics, Ecole Polytechnique 
F\'ed\'erale de Lausanne (EPFL), Observatoire de Sauverny, 1290 Versoix, 
Switzerland \label{epfl}
\goodbreak
\and
Department of Physics and Astronomy, Lehman College of the CUNY, Bronx, NY 10468, USA \label{lehman}
\goodbreak
\and
American Museum of Natural History, Department of Astrophysics, New York, NY 10024, USA \label{amnh}
\goodbreak
\and
Borough of Manhattan Community College, City University of New York,  Department of Science, New York, NY 10007, USA \label{bmcc}
\goodbreak
\and
STAR Institute, University of Li{\`e}ge, Quartier Agora, All\'ee du six Ao\^ut 19c, 4000 Li\`ege, Belgium \label{uliege}
}

\abstract{Challenges inherent to high-resolution and high signal-to-noise data as well as model degeneracies can cause systematic biases in analyses of strong lens systems. In the past decade, the number of lens modeling methods has significantly increased, from purely analytical methods, to pixelated and non-parametric ones, or ones based on deep learning. We embraced this diversity by selecting different software packages and use them to blindly model independently simulated \textit{Hubble} Space Telescope (HST) imaging data. To overcome the difficulties arising from using different codes and conventions, we used the COde-independent Organized LEns STandard (COOLEST) to store, compare, and release all models in a self-consistent and human-readable manner. From an ensemble of six modeling methods, we studied the recovery of the lens potential parameters and properties of the reconstructed source. In particular, we simulated and inferred parameters of an elliptical power-law mass distribution embedded in a shear field for the lens, while each modeling method reconstructs the source differently. We find that, overall, both lens and source properties are recovered reasonably well, but systematic biases arise in all methods. Interestingly, we do not observe that a single method is significantly more accurate than others, and the amount of bias largely depends on the specific lens or source property of interest. By combining posterior distributions from individual methods using equal weights, the maximal systematic biases on lens model parameters inferred from individual models are reduced by a factor of 5.4 on average. We investigated a selection of modeling effects that partly explain the observed biases, such as the cuspy nature of the background source and the accuracy of the point spread function. This work introduces, for the first time, a generic framework to compare and ease the combination of models obtained from different codes and methods, which will be key to retain accuracy in future strong lensing analyses.}


\maketitle


\section{Introduction}

Understanding the evolutionary path of galaxies over cosmic times continues to be a major challenge in astrophysics. In this context, strong gravitational lensing enables the observation of galaxies lying at different redshifts in a single observation, making it an inescapable tool to constrain galaxy evolution models. Strong gravitational lensing arises when a foreground distant galaxy -- the lens, or deflector -- is coincidentally aligned with a more distant background galaxy -- the source -- causing the appearance of multiple and magnified images of the latter. The typical redshift range for lens galaxies lies between $z_{\rm d}\sim0.2$ and $1.5$, while source galaxies are often found between redshifts $z_{\rm s}\sim1$ and $4$ \citep{OguriMarshall2010,Collett2015}, such that strong lensing systems can display a wide variety of galaxy morphologies and evolutionary stages.

Besides galaxy evolution studies, strong lensing has several important applications in cosmology. As it is dictated by the total mass of galaxies, one can use this effect to put constraints on their dark matter halo. In particular, strong lensing data enables the separation of the baryonic and dark components of galaxies \citep[e.g.,][]{Suyu2012glee,Shajib2021}, and the detection of dark matter subhalos and other invisible masses along the line of sight \citep[e.g.,][]{Vegetti2010,Sengul2022,Nightingale2024}. When combined with time-varying sources, strong lenses can also be used to measure cosmological parameters, including the \textit{Hubble} constant \citep[$H_0$ e.g.,][]{Wong2020,Birrer2020,Kelly2023} and density parameters \citep[e.g., with multiplane lensing systems,][]{Collett2014}. All these applications rely heavily on a precise characterization of both the azimuthal and radial mass profiles of lens galaxies.

The central step when analyzing strong lensing data is lens modeling. The goal of this step is to model both the mass and light distribution of the lens galaxy, while simultaneously reconstructing an unlensed version of the source galaxy. Lens modeling is a challenging task because inverting the lensing effect is an ill-posed problem, in particular due to known degeneracies between the lens mass distribution and the source morphology. For example, the infamous mass-sheet degeneracy \citep[MSD,][]{Falco1985,SchneiderSluse2013}, a mathematically exact degeneracy between the lens mass density and the source scaling, has been studied both theoretically and practically \citep[e.g.,][]{Birrer2016,Unruh2017,Wagner2018,GomerWilliams2020,Cao2022} and can be mitigated using complementary data sets \citep[e.g.,][]{Birrer2020,Yildirim2023,Khadka2024}. In the past twenty years, many different lens modeling techniques, ranging from analytical to pixelated techniques and neural networks, have been developed and successfully applied to real images \citep[e.g.,][]{ WarrenDye2003,Suyu2006,VegettiKoopmans2009,Birrer2016,Nightingale2018,Galan2024corrfield}. In general, these techniques have been developed with specific lensing systems, data sets, and science goals in mind, and then have been extended to cover more use cases. Consequently, it is crucial to assess how these different methods compare to each other, and if their combination is warranted to improve the robustness of lensing analyses. Such a comparison enables the quantification of possible systematic biases. Additionally, if several methods lead to consistent results, those can be combined together, improving the overall accuracy.

So far, lens modeling comparison analyses have been rare. For cluster-scale systems, a prominent work has been initiated by \citet{Meneghetti2017} by performing an extensive comparison of several modeling approaches on both simulated and real \textit{Hubble} Frontier Fields clusters. However, there have not been comparable efforts for galaxy-scale strong lens systems. While some works have reanalyzed archival data with alternative modeling software packages \citep[e.g.,][]{Birrer2016,Shajib2021}, it is only recently that more quantitative comparisons between different methods have been reported \citep{ShajibWong2022,Etherington2022}. The Time Delay Lens Modeling Challenge (TDLMC) compared the output of different modeling and inference strategies, but focusing only on the recovery of $H_0$ \citep{Ding2021}. Different lens modeling codes have been compared in \citet{LeforFutamase2015} and \citet{Pascale2024}, although using point-like multiple images rather than extended gravitational arcs as constraints. Finally, recent works from \citet{Schuldt2023comp} and \citet{Gawade2024} compared neural network predictions with more classical approaches, although on with ground-based imaging data.

Our goal here is to analyze imaging data similar to those obtained with the \textit{Hubble} Space Telescope (HST) with different modeling and inference techniques, and study the recovery of a given set of lens parameters while reconstructing the lensed source in different ways. To our knowledge, this is the first time a systematic and self-consistent comparison between a large number (six) of state-of-the-art galaxy-scale lens modeling methods has been conducted. In order to maintain this novel kind of analysis tractable, we have restricted the assumptions regarding the description of the lens mass distribution and properties of the data, although remaining reasonably realistic. In particular, we limit our scope to the commonly used power-law elliptical mass distribution embedded in a shear field. This description of the lens deflection field has proven to be a minimal but efficient prescription for modeling the observed strong lensing effect caused by large elliptical galaxies \citep[e.g.,][to cite only a few]{Koopmans2006,Suyu2013,Millon2020,Shajib2021,Etherington2022,Tan2024}, although the simplicity of this model has known limitations \citep[][]{Sonnenfeld2018,GomerWilliams2021,Cao2022,Etherington2024shear,RuanKeeton2023}. We also note that recent analyses of strong lenses found evidence for multipolar deviations to the elliptical power-law profile, but a higher resolution than HST is warranted for robust detection \citep{Powell2022,Stacey2024}. While most lensing analyses focus on the properties of the galaxies acting as lenses, the morphology of the lensed galaxies also hold important information about galaxy formation and evolution. Current high-resolution images of strong lenses such as those from HST showcase highly structured lensed sources \citep[e.g.,][]{Bolton2006,Garvin2022,Wang2022}. Consequently, it is also crucial to assess the ability of lens modeling codes to recover the morphology of extended lensed sources.

We first selected different lens modeling software packages and modeling methods that are well suited to model high-resolution and high signal-to-noise (S/N) data. Since each software package typically follows different parameter definitions and model conventions, it is not possible to directly compare the modeling results. We overcame this challenge by using the COde-independent Organized LEns STandard \citep[\coolest,][]{Galan2023coolest}, an open-source standard that enables storage, sharing, and analysis of all lens modeling products in a uniform manner, regardless of the modeling code originally used to perform the lens modeling tasks. Included in this standard is an analysis interface allowing us to compute important quantities (e.g., effective radii and profile slopes) and visualize lens modeling results. We have extensively used \coolest in this work, both for releasing the models and data, as well as performing the analysis of the results and generating the figures.

The paper is organized as follows. In Sect.~\ref{sec:lensing_formalism} we briefly recall the strong lensing formalism we follow. In Sect.~\ref{sec:modeling_methods}, we present the different lens modeling methods, in particular their commonalities and differences. We explain how the data was simulated using an independent software in Sect.~\ref{sec:data_simulation}, and the standardized comparison and analysis framework is introduced in Sect.~\ref{sec:coolest}. The modeling results after unblinding are visualized and described in Sect.~\ref{sec:results_mock_data}, followed by an exploration of possible sources of systematics in Sect.~\ref{sec:systematics_tests}. In Sect.~\ref{sec:discussion} we discuss our results and place them in a broader context, and Sect.~\ref{sec:conclusion} concludes our work.

\section{Formalism  of strong gravitational lensing \label{sec:lensing_formalism}}

We give for completeness a brief overview of the mathematical formalism to describe strong lensing data and models. More background details can be found in recent reviews such as \citet{Vegetti2023review,Shajib2024review,Saha2024review}.

The main strong lensing observables are the positions and intensities of multiply lensed images of features in a background sources. These features can either be unresolved (i.e., point sources) or spatially extended. In the latter case, the lensed source appears as several arcs or as an Einstein ring surrounding the lens object and typically covering many pixels in high-resolution imaging data. We call the (observable) plane of the sky where lensed images appear the ``image plane'' , that we place at the redshift $z_{\rm d}$ of the foreground lens, also called the main deflector. Observed features in the image plane are localized with a two-dimensional angular position vector, $\boldsymbol{\theta}$. For conciseness, we interchangeably use the standard Cartesian coordinates $(x, y)$ to describe a position, $\boldsymbol{\theta}$, in the image plane. Each feature in the lensed images has a corresponding (unobservable) angular position, $\boldsymbol{\beta}$, in the ``source plane'' placed at the redshift of the background object, $z_{\rm s}$.

The central equation in gravitational lensing is the lens equation, which gives the relationship between $\boldsymbol{\beta}$ and $\boldsymbol{\theta}$:
\begin{align}
    \label{eq:lens_equation}
    \boldsymbol{\beta} = \boldsymbol{\theta} - \nabla\psi(\boldsymbol{\theta}) \ ,
\end{align}
where $\nabla\psi\equiv\boldsymbol{\alpha}$ is the deflection field originating from the lens potential, $\psi$, the latter being a rescaled and projected version of the underlying three-dimensional gravitational potential of the lens galaxy. Usually, a more physically relevant quantity is the projected mass density of the lens, characterized by the so-called lens convergence, $\kappa$ (dimensionless), obtained with a combination of second derivatives of the lens potential:
\begin{align}
    \kappa = \frac12 \nabla^2\psi \ .
\end{align}

As it will be useful for the discussion (Sect.~\ref{ssec:time_delay_app}), we also recall the formula of the Fermat potential, mostly relevant for time-varying sources. The Fermat potential, $\Phi_i$, and the Fermat potential difference, $\Delta\Phi_{ij}$, between a pair of lensed images $i$ and $j$, are defined as
\begin{align}
    \label{eq:fermat_pot}
    \Phi(\boldsymbol{\theta}_i) &= \frac{\big(\boldsymbol{\theta}_i - \boldsymbol{\beta}\big)^2}{2} + \psi(\boldsymbol{\theta}_i) \ ,\\
    \label{eq:fermat_pot_diff}
    \Delta\Phi_{ij} &\equiv \Phi(\boldsymbol{\theta}_i) - \Phi(\boldsymbol{\theta}_j) \ ,
\end{align}
where $\boldsymbol{\theta}_i$ and $\boldsymbol{\theta}_j$ are the positions of images $i$ and $j$, respectively.

In this work, we consider parametrized forms for the lens mass distribution, while the surface brightness of the lensed galaxy is described following a variety of techniques. We give more details about the modeling of these different components in Sect.~\ref{sec:modeling_methods}.

\section{Lens modeling methods and assumptions \label{sec:modeling_methods}}

We have considered an ensemble of ``modeling methods'' that each differ on two aspects: modeling assumptions and inference techniques. For instance, modeling assumptions are typically specific choices of model components (mass and light profiles, fixed or not), regularization strategies for pixelated models and necessary hyper-parameters. Inference techniques are typically minimization and sampling algorithms to obtain best-fit parameters and estimate their posterior distributions, or sequences of distinct steps (e.g., preliminary coarse and fast model fits) to converge to the best-fit solution.

In practice, a given lens modeling software package can be considered as a modeling method, as specific choices regarding the code structure, model types and optimization techniques have been made throughout its development. For this work, we selected a subset of software packages that are sufficiently different to be considered as distinct modeling methods: \lenstronomy \citep{Birrer2018lenstro,Birrer2021lenstro}, the Very Knotty Lenser \citep[\vkl,][]{Vernardos2022vkl}, \herculens \citep{Galan2022herculens,Galan2024corrfield} and \qlens (Minor et al., in prep.). Other software packages used in several published analyses so far are \glee \citep{Suyu2010glee,Suyu2012glee}, \textsc{PyAutoLens} \citep{Nightingale2015,Nightingale2018,pyautolens}, \textsc{glafic} \citep{Oguri2010} and methods from \citet[and subsequent works]{VegettiKoopmans2009}. However, for practical reasons we only use the first set of methods, which already form a representative sample of the various modeling methods that are currently available, from fully analytical to pixelated models, with or without adaptive grids. Such methods are referred to as classical methods, in contrast to  deep learning methods that we do not consider in this work \citep[e.g.,][for some recent works]{Schuldt2023,Adam2023,Gentile2023}, as these would require additional assumptions regarding training sets and network architectures beyond our scope. Nevertheless, we encourage future works to conduct self-consistent comparison analyses similar to ours, that involve both classical and deep learning methods \citep[see e.g.,][]{Schuldt2023comp}.

This remaining of this section presents the general modeling strategy we adopt throughout this work. We first describe modeling assumptions that are common to all methods, then give more details regarding each of these modeling methods, and finally mention extra choices that are left free to the modelers.

\subsection{Common modeling aspects \label{ssec:common_aspets}}

Throughout this work, we reasonably assume that the noise in the imaging data, $\boldsymbol{d}$, follows a Gaussian distribution with covariance matrix, $C_d$. In this setting, we can write the negative log-probability of the data likelihood as
\begin{align}
\label{eq:data_likelihood}
    \nonumber
    - \log\,\mathcal{L}\big(\boldsymbol{\eta}\big) 
    = & \ \frac12\,\bigg[\boldsymbol{m}(\boldsymbol{\eta}) - \boldsymbol{d}\bigg]^\top C_d^{-1}\, \bigg[\boldsymbol{m}(\boldsymbol{\eta}) - \boldsymbol{d}\bigg] \\
    &+ \log\bigg( 2\pi\sqrt{\det C_d}\bigg) \ ,
\end{align}
where $\boldsymbol{m}$ is the predicted image (i.e., the model) and $\boldsymbol{\eta}$ represents a generic vector of model parameters. We note that $C_d$ is assumed to be diagonal with contributions from both background noise and photon noise. In other words, we follow the widely used assumption that the noise is uncorrelated and normally distributed. With simulated data, we have access to the true matrix $C_d$. As in this work we do not explore the effects of inaccurate assumptions regarding noise characteristics, we give to the modelers the true matrix $C_d$ and use it in all lens models.

While the likelihood term in Eq.~\ref{eq:data_likelihood} is common to all models considered here, specific modeling assumptions such as morphological properties of the source galaxy are encoded as additional priors. Such priors priors can either be explicitly incorporated in the inference via a regularization term written as the negative of the log-prior $-\log\mathcal{P}$, or they can be implicitly defined through a choice of parametrization such as an analytical functions. Summing the log-likelihood and log-prior terms gives the full penalty or loss function, $L$, which is directly proportional to the log-posterior and minimized during the inference of model parameters:
\begin{align}
\label{eq:loss_func}
    L\big(\boldsymbol{\eta}\big) = -\log\mathcal{L}\big(\boldsymbol{\eta}\big) - \log\mathcal{P}\big(\boldsymbol{\eta}\big) \ .
\end{align}

Modeling methods generally describe the lensing of photons from the source by casting the lens equation into a lensing operator, $\boldsymbol{\mathsf{L}}$, which depends on the lens potential parameters that we denote by $\boldsymbol{\eta_{\rm \psi}}$. This operator acts on a model of the source, $\boldsymbol{s}$, described by parameters, $\boldsymbol{\eta_{\rm s}}$, which can be either analytical, pixelated or a representation in function basis set, as per 
\begin{align}
\label{eq:image_model}
    \boldsymbol{m}(\boldsymbol{\eta}) \equiv \boldsymbol{m}(\boldsymbol{\eta}_{\psi},\boldsymbol{\eta}_{\rm s}) = \boldsymbol{\mathsf{R}}\, \boldsymbol{\mathsf{B}}\, \boldsymbol{\mathsf{L}}(\boldsymbol{\eta}_\psi)\, \boldsymbol{s}(\boldsymbol{\eta}_{\rm s}) \ .
\end{align}
so that we get a model image, $\boldsymbol{m}$, that has the same pixel size as the data, after possible downsampling by the operator $\boldsymbol{\mathsf{R}}$ and blurring by the operator $\boldsymbol{\mathsf{B}}$. The latter incorporates the effect of the point spread function (PSF) of the instrument and seeing conditions. This PSF is assumed to be known with the same spatial sampling as the data and available to all modelers. As stated in Sect.~\ref{ssec:modeling_freedom}, no constraints are imposed to modelers regarding  optional supersampling of ray-tracing and convolution operations. The light distribution of the lens galaxy is not modeled because we assume that the lens light has been perfectly subtracted from the data beforehand.

For modeling the lens mass distribution of the lensing galaxy---alternatively, its lens potential---we consider the commonly used power-law elliptical mass distribution (PEMD) with an additional shear field component. This shear component has been commonly referred to as ``external'' shear as it captures the net effect masses external to the lens along the line of sight; we use this wording within the scope of our work but it does not necessarily holds true when modeling real systems \citep[see e.g.,][]{Etherington2024shear}.

The assumption of a PEMD with external shear is common to all modeling methods, namely all modelers use the same lens potential parameter vector $\boldsymbol{\eta_{\rm \psi}}$. The convergence of the PEMD is described by \citep{Barkana1998,Tessore2015}:
\begin{align}
    \label{eq:def_pemd}
    \kappa_{\rm PEMD}(x,y) = \frac{3-\gamma}{2} \left(\frac{\theta_{\rm E}}{\sqrt{q_{\rm m}x^2+y^2/q_{\rm m}}} \right)^{\gamma -1}
\end{align}
where $\gamma$ is the logarithmic power-law slope ($\gamma=2$ corresponding to an isothermal profile), $q_{\rm m}$ is the axis ratio and the coordinate system $(x,y)$ has been rotated by a position angle $\phi_{\rm m}$ around the lens center $(x_0,y_0)$. The lens potential generated by an external shear can be easily expressed in polar coordinates with the following formula \citep[e.g.,][]{Etherington2024shear}:
\begin{align}
    \label{eq:def_extshear}
    \psi_{\rm ext}(x, y) \equiv \psi_{\rm ext}(r,\phi) = \frac{r^2}{2} \, \gamma_{\rm ext} \, \cos\bigg[ 2 \big(\phi - \phi_{\rm ext}\big) \bigg],
\end{align}
where $\gamma_{\rm ext}$ is the strength of the external shear, and $\phi_{\rm ext}$ its position angle. We note that Eqs.~\ref{eq:def_pemd} and \ref{eq:def_extshear} follow the parameters conventions used in the \coolest (see Sect.~\ref{sec:coolest} and the online documentation for other conventions\footnote{\url{https://coolest.readthedocs.io/en/latest/conventions.html}}).

We note that the loss function of Eq.~\ref{eq:loss_func} has a non-linear response to lens mass parameters, $\boldsymbol{\eta}_\psi$. However, the set of source parameters, $\boldsymbol{\eta}_{\rm s}$, can formally be split into linear (light profile amplitudes) and non-linear parameters. This property is explicitly exploited by some of the modeling methods we use in this work.

\begin{table*}[h!]
    \caption{Labels used to identify the six modeling methods of this work. These modeling methods mainly differ in their source modeling techniques, but share other differences (see Sect.~\ref{sec:modeling_methods} for details). \label{tab:modeling_methods}}
    \renewcommand{\arraystretch}{1.5}
    \centering
    \small
    \begin{tabular}{lll}
         \normalsize{Modeling method label} & \normalsize{Description} & \normalsize{Corresponding source modeling technique} \\
         \hline\hline
         \normalsize{Sérsic+Shapelets} & Sect.~\ref{ssec:general_lenstronomy} & Concentric elliptical Sérsic profile and shapelets basis functions \\ 
         \normalsize{Adaptive+Matérn} & Sect.~\ref{ssec:general_semlinear_inv} & Adaptive Voronoi grid, regularized with a Matérn kernel \\
         \normalsize{Cluster+Exp} & Sect.~\ref{ssec:general_semlinear_inv} & Clustered adaptive Voronoi grid, regularized with an exponential kernel \\
         \normalsize{Cluster+Exp+Lumweight} & Sect.~\ref{ssec:general_semlinear_inv} & Similar to Cluster+Exp with regularization weighted by source luminosity \\
         \normalsize{Sparsity+Wavelets} & Sect.~\ref{ssec:general_multiscale} & Cartesian grid, regularized with sparsity in wavelets space and non-negativity constraints \\
         \normalsize{Correlated Field} & Sect.~\ref{ssec:general_fields} & Cartesian grid with Gaussian processes, regularized with a parametric power-spectrum \\
    \end{tabular}
\end{table*}

The modeling methods considered in this work thus mainly differ in the assumptions regarding the light distribution of the source, $\boldsymbol{s}(\boldsymbol{\eta}_{\rm s})$, which we summarize in Table~\ref{tab:modeling_methods} and describe in more detail in the next subsections. %


\subsection{Smooth modeling with Sérsic and shapelets \label{ssec:general_lenstronomy}}

We used a modeling method implemented in the multipurpose \lenstronomy package. We followed the baseline model presented in Sect.~\ref{ssec:common_aspets}, and which consists of a PEMD plus external shear for the lens. The source was modeled with a Sérsic profile, to which were added shapelets basis functions, which are capture additional complexity of the source light distribution. We implemented this modeling strategy using version 1.11.3 of the multi-purpose open-source software package \lenstronomy~\citep{Birrer2018lenstro,Birrer2021lenstro}. This tool, regularly enhanced with new user-contributed capabilities, provides a large family of lens mass distribution and light profiles. We refer the reader to \citet{Birrer2015} for a formal description of the shapelets model in the context of lens modeling.

The procedure used in \lenstronomy to derive the posterior distribution on the parameters is sequential. The optimal linear parameters are found through matrix inversion, given values of non-linear parameters. First, a suitable region in non-linear parameter space that minimizes the loss function defined in Eq.~\ref{eq:loss_func} is found via a Particle Swarm Optimization algorithm \citep[PSO,][]{KennedyEberhart2002_pso}. Second, the parameters space is sampled using a Monte-Carlo Markov Chain (MCMC). The parameters of the optimal model found previously are randomly perturbed, and used to start the chain. We use the MCMC sampler \textsc{emcee}, which is the most used so far among the \lenstronomy user community \citep{ForemanMackey2013_emcee}. In this work, as it is also a common practice, the model investigated during the optimization step is a simplified version of the final model, retaining only the main model components that enable to reproduce the largest fraction of the data pixel values. Components that yield small changes of the loss function, such as the source shapelets, are added only during the MCMC sampling. This hierarchy in the significance of model parameters, while not explicitly formalized in the code, is similar to the methodology developed by several automatized lens modeling efforts \citep[e.g.,][]{Etherington2022,Ertl2023,Tan2024}. We give more technical details regarding this method in Sect.~\ref{app:ssec:sersic_shapelets}.

\subsection{Adaptive grid source modeling \label{ssec:general_semlinear_inv}}

If one assumes that the free parameters of the source are its brightness values cast on a grid of pixels $\boldsymbol{s}$ (instead of being defined from a continuous analytical profile), then the likelihood of Eq.~\ref{eq:data_likelihood} becomes a quadratic function of $\boldsymbol{s}$. The benefit of such quadratic functions is that their derivative can be calculated analytically and have a unique minimum \citep[][]{WarrenDye2003}. However, with just the likelihood term this leads to an ill-posed problem and the addition of a (quadratic) regularization term is required, which has the following generic form:
\begin{equation}
    \label{eq:quadratic_reg_general}
    -\log\mathcal{P}(\lambda,\boldsymbol{g},\boldsymbol{s}) =  \frac{1}{2} \lambda \boldsymbol{s}^T C_s^{-1}(\boldsymbol{g}) \boldsymbol{s} \ ,
\end{equation}
where $C_s$ is some covariance kernel of the source as a function of parameters, $\boldsymbol{g}$.
In this form, the source parameters can be obtained analytically from $\nabla_s L = 0$ once the lens parameters, $\boldsymbol{\eta}_\psi$, regularization strength, $\lambda$, and covariance parameters, $\boldsymbol{g}$, are given.
This approach is referred to as semi-linear inversion.
In comparison with forward methods, which may treat more source parameters as non-linear parameters, there are only a few additional parameters that require sampling, $\lambda$ and $\boldsymbol{g}$ (usually corresponding to just one or two parameters). The linear source parameters, in other words the pixel brightness values, are obtained using matrix inversion.

A key assumption in such inverse problems is the choice of regularization, which can be interpreted in a Bayesian way as a prior imposed on the source.
Traditionally, one may choose to impose smoothness to the solution through its derivatives, where the matrix $C_{s}^{-1}$ is constructed from the numerical derivative coefficients computed on the pixelated grid \citep{WarrenDye2003,Suyu2006,VegettiKoopmans2009}.
Alternatively, more physically motivated covariance kernels obtained from real galaxy brightness distributions have been shown to perform better and lead to less biased results \citep{Vernardos2022vkl}.
A quite generic such example is the Matérn kernel that has the following analytic form:
\begin{equation}
    \label{eq:mattern}
    C(r_{ij}|l,\nu) = \frac{2^{1-\nu}}{\Gamma(\nu)} \left( \frac{r_{ij}\sqrt{2\nu}}{l} \right)^\nu K_{\nu} \left( \frac{r_{ij} \sqrt{2\nu}}{l} \right),
\end{equation}
where $r_{ij}$ is the distance between any two source pixels and $\nu,l$ correspond to the non-linear parameters $\boldsymbol{g}$ (the latter can be interpreted as a correlation length).
The type of regularization can be objectively chosen based on the Bayesian evidence.

Another feature employed in semi-linear inversion implementations is the use of adaptive, non-regular grids for the source.
This is because smaller regions in the source plane can contribute a higher fraction of the total flux, especially in the regions of high magnification near caustics, hence higher resolution is required.
Using a high resolution fixed regular grid will lead to more computationally demanding matrix inversions, but adapting the resolution to the lensing magnification (which is a function of the lens model parameters $\boldsymbol{\eta}$) will increase the resolution in those regions of the source plane where it is needed without adding more degrees of freedom.
Such adaptive grids can be constructed simply by tracing a subset of the data pixels back to the source \citep{VegettiKoopmans2009,Vernardos2022vkl} and using them as grid vertices. A more sophisticated way of constructing the adaptive grid, pioneered by \citet{Nightingale2015}, is to split each pixel into $N\times N$ subpixels, ray-trace the subpixels to the source plane, and then use a $k$-means clustering algorithm to determine the location of the source grid vertices. The latter approach has the advantage of minimizing aliasing effects, at the cost of additional overhead due to the clustering algorithm.

The regularization of an adaptive grid may be extended to allow for greater fluctuations in surface brightness in the inner regions of the source with high S/N, while keeping the outer regions of the source relatively smooth. Luminosity-weighted regularization has been explored by \citet{Nightingale2018} in the context of gradient regularization. In this work, we explored luminosity-weighted covariance kernels to achieve a similar effect. To implement this, we define the luminosity-weighted kernel as:
\begin{equation}
    \label{eq:cij_lum}
    C_{ij,lum} = C_{ij} W_i W_j,
\end{equation}
with the weighting function $W_i$ given by:
\begin{equation}
    \label{eq:weightfunc}
W_i = \exp \left[ -\rho \, (1-s_{i,0}/s_{\rm max}) \right],
\end{equation}
where $\rho$ is a free parameter to be varied, $s_{i,0}$ is an approximate surface brightness of the $i$-th pixel before luminosity weighting, and $s_{\rm max}$ is the maximum surface brightness of all the source pixels. Pixel values $s_{i,0}$ can be taken from the best-fit model of a previous fit altogether, or it can be estimated during each likelihood evaluation by doing an inversion without luminosity weighting first; here, we adopted the latter approach. The advantage of the form given in Eq.~\ref{eq:cij_lum} is that it ensures that the kernel remains positive-definite, which is critical given the quadratic form of the regularization term. Although more sophisticated forms of the weighting function $W_i$ are possible, we only explored the single parameter function given by Eq.~\ref{eq:weightfunc} in this work.

Here, we used two implementations of the semi-linear inversion technique, \vkl \citep{Vernardos2022vkl} and \qlens (Minor et al., in prep.), which can handle different regularization schemes and recipes for constructing the adaptive source-pixel grid. In the \vkl modeling run, which we label Adaptive+Matérn, a Matérn kernel will be used for regularization without supersampling of the image plane; whereas in the \qlens modeling runs, an exponential kernel will be used (equivalent to Matérn with $\nu=0.5$), image plane supersampling with $k$-means clustering will be used to construct the adaptive source grid, and models without and with luminosity weighted regularization will be used (Cluster+Exp and Cluster+Exp+Lumweight, respectively).
Both implementations use Nested Sampling \citep[as implemented by the MultiNest algorithm,][]{Feroz2009_multinest} to sample the parameter space and converge to the maximum a posteriori solution, in addition to calculating the Bayesian evidence. 
We outline the specific modeling choices that are considered as three different source modeling techniques in Sects~\ref{app:ssec:ag_vor_mat} and \ref{app:ssec:ag_vor_clust}.

\subsection{Multi-scale regularization with wavelets \label{ssec:general_multiscale}}

Regularization does not necessarily have to be quadratic so that it can lead to a linear solution for the source. A non-linear example is the method by \citet{Joseph2019,Galan2021} that is based on sparsity constraints in the wavelet domain. We refer to this regularization as a multi-scale regularization (and use ``ms'' as subscripts in the corresponding equations). We briefly describe the method here, but refer the reader to the original papers for the full mathematical treatment.

Given the vector representation of the source expressed on a regular (non-adaptive) pixelated grid, the regularization term to be minimized (i.e., the second term in Eq.~\ref{eq:loss_func}) is:
\begin{align}
\label{eq:wavelet_regul}
    - {\rm log}\,\mathcal{P}(\boldsymbol{\eta},\boldsymbol{s},\lambda_{\rm ms}) = \lambda_{\rm ms} \big\| {\rm \bf W_{\rm ms}}(\boldsymbol{\eta_{\rm \psi}}, \boldsymbol{\eta_{\rm s}}) \odot \boldsymbol{\mathsf{\Phi}}^\top\,\boldsymbol{s} \big\|_1 + i_{\geq0}(\boldsymbol{s}) \ ,
\end{align}
where $\lambda_{\rm ms}$ is a global (scalar) regularization parameter, $\boldsymbol{\mathsf{\Phi}}^\top$ is the wavelet transform operator that transforms $\boldsymbol{s}$ into its wavelets coefficients, and ${\rm \bf W_{\rm ms}}$ is a matrix that scales the regularization strength for each these coefficients. Operations $\|\cdot\|_1$ and $\odot$ refer to the $\ell_1$ norm and element-wise product, respectively. The second term in Eq.~\ref{eq:wavelet_regul} is a non-negativity constraint on the source since we reconstruct surface brightness values. The operator $\boldsymbol{\mathsf{\Phi}}^\top$ represents an hybrid wavelet transform composed of all scales from of the starlet transform \citep{Starck2007} and the first scale of the Battle-Lemarié wavelet transform \citep[for more details see e.g.,][]{Lanusse2016,Galan2022herculens}. The matrix elements of ${\rm \bf W_{\rm ms}}$ are computed by propagating the data noise from the source plane to the wavelet domain (hence the dependence on $\boldsymbol{\eta_{\rm \psi}},\boldsymbol{\eta_{\rm s}}$), allowing us to attach a clear meaning to the global regularization strength $\lambda_{\rm ms}$, interpreted as the statistical significance of the regularized source model and given in units of the noise (e.g., $\lambda_{\rm ms} = 3\sigma$).

We dynamically adapt the extent of the source plane regular grid based on the lens mass by defining the annular region in image plane (the ``arc mask'') within which the pixelated source light is evaluated. This adaptive scheme ensures that the effective source pixel size covers the same angular scale of the source for any realization of the mass model\footnote{We note that a similar strategy is used in the \glee modeling code to adapt the extent of the source plane regular grid \citep{Suyu2010glee,Suyu2012glee}}. This treatment is different from the fixed source plane grid used in \citet{Galan2021}, and provides more stability when optimizing lens mass models with a free density slope.

We used the strong lens modeling code \herculens~\citep{Galan2022herculens} that implements the multi-scale regularization strategy described in Eq.~\ref{eq:wavelet_regul} using JAX \citep{jax2018github}, such that the full model can be pre-compiled and is differentiable. We used the probabilistic programming library \textsc{NumPyro} \citep{phan2019composable,bingham2019pyro} to implement prior distributions and constraints on model parameters, and to perform the inference of posterior distributions. Additional technical details regarding the Sparsity+Wavelets models are given in Sect.~\ref{app:ssec:ms_wavelets}.

\subsection{Non-parametric Gaussian processes \label{ssec:general_fields}}

Finally, we applied a recently introduced source reconstruction method that relies on Gaussian processes and information field theory \citep[IFT][]{Ensslin2019}. The first applications of Gaussian processes for modeling strongly lensed sources are the recent works of \citet{Karchev2022,Rustig2024_lenscharm} and \citet{Galan2024corrfield}, but we describe below the main principles for completeness. In the IFT framework, such Gaussian processes are often referred to as correlated fields, which is the term we use in the remaining.

We use IFT to represent the source light distribution as a two-dimensional non-negative correlated field. Such a field is based on two main components: (1) an analytical parametrization of its power spectrum (mainly its amplitude and slope) that accounts for correlated structures in the source; (2) a discretization onto a regular square grid, with elements following standardized Gaussian distributions (we call the latter an excitation field). More formally, we describe the source field, $\boldsymbol{s}$, as
\begin{align}
\label{eq:src_correl_field}
    \boldsymbol{s} = \exp\bigg[ \boldsymbol{\mathsf{F}}^{-1} \big( \boldsymbol{A}_0 \odot \boldsymbol{\xi} \big) + \boldsymbol{\delta} \bigg] \ ,
\end{align}
where $\boldsymbol{A}_0$ is a zero-mode spectral field generated from the parametrized power spectrum, $\boldsymbol{\xi}$ is the excitation field, and $\odot$ is the point-wise multiplication. The resulting field in harmonic space is then transformed to real space by applying the inverse Fourier transform operator, $\boldsymbol{\mathsf{F}}^{-1}$, to which a constant offset, $\boldsymbol{\delta}$, is added. Finally, we take the exponential of the resulting field to enforce the positivity of source pixel values, since they represent surface brightness. We note that no extra regularization term is added to the loss function, as Eq.~\ref{eq:src_correl_field} describes a generative model that already incorporates smoothness conditions through its power spectrum.

We use the Python library \nifty\footnote{\url{https://gitlab.mpcdf.mpg.de/ift/nifty}} \citep{nifty1,nifty3,nifty5} to implement the above field model and variational inference samplers to get the joint posterior distribution over the parameter space. As in \citet{Galan2024corrfield}, we use the JAX interface of \nifty \citep[\texttt{nifty.re},][]{niftyre} and combine it with \herculens to evaluate the forward model (Eq.~\ref{eq:image_model}). For further technical details regarding the Correlated field model, see Sect.~\ref{app:ssec:correl_field}.

\subsection{Additional choices left to the modelers \label{ssec:modeling_freedom}}

There are some additional choices that are left to the modelers. In particular, they are free to choose if and how they mask out some regions in the imaging data and exclude those from the data likelihood evaluation. Similarly, super-sampling of the coordinates grid when performing ray-tracing evaluations and surface brightness convolutions with the PSF are optional. Modelers are free to run variations in their fiducial models, varying, for example, hyper-parameters, random number generator seeds, or inference algorithms. This may allow for robustness tests, and any eventual marginalization over families of models leading to a joint posterior distribution as the final result.

\section{Independent simulation of imaging data \label{sec:data_simulation}}

We produced a simulated mock system to model with all methods presented above and compare the results. We simulated the mock using \molet\footnote{\url{https://github.com/gvernard/molet}} \citep{Vernardos2022molet}, a simulator code that is independent of any of the codes used to fit the mock. The simulated mock is constructed using a power-law mass profile with external shear and a relatively structured source, as is expected from most lensed galaxies. The true source used in the simulation is kept hidden (blind) from the modelers, as well as all other input parameters.

For the source, we used an HST image of the galaxy NGC\,1084 that was previously used for simulated lenses in the context of the TDLMC \citep{Ding2021}. This source is a local galaxy with a detailed structure that is resolved without any prominent PSF spikes, which could introduce nonphysical features if lensed. We further avoided introducing unphysical features in the resulting lensed source due to edge effects (e.g., sky background) in the cutout that we used by forcing the values in the source pixels to decay exponentially to zero away from the brightness peak.

\begin{figure}
    \centering
    \includegraphics[width=\linewidth]{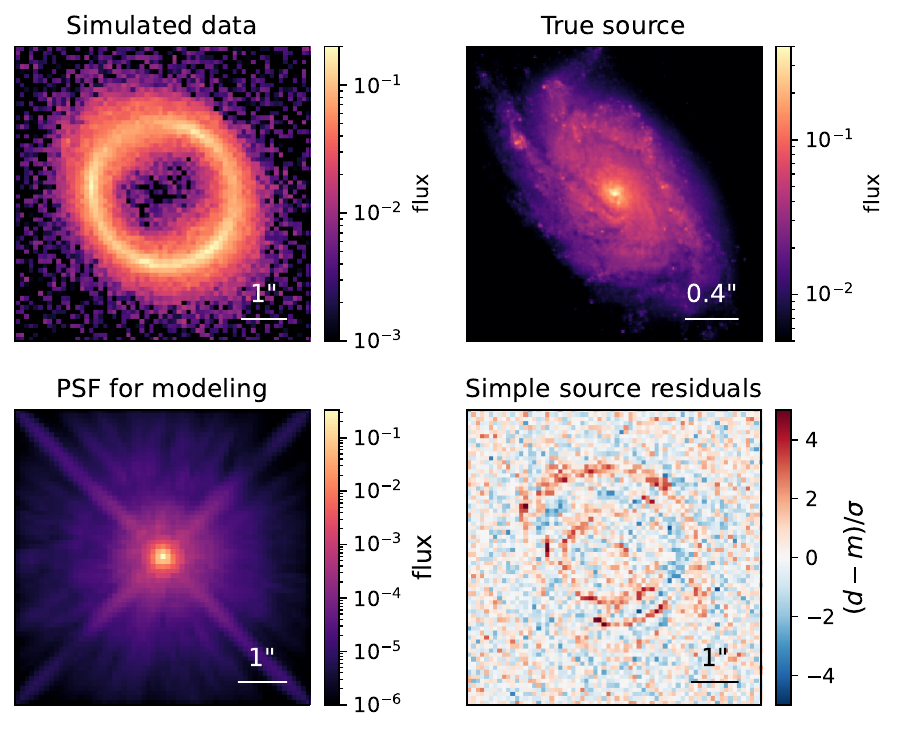}
    \caption{Simulated HST imaging data used for the blind lens modeling experiment (see Sects.~\ref{sec:data_simulation} for details). The top left panel shows a zoom-in cutout of the data. The top right panels shows the true (unlensed) source surface brightness. The bottom left panel shows the true PSF kernel downscaled to the data resolution (and given to the modelers). The bottom right panel shows normalized residuals in image plane obtained with a too simplistic source model (a single Sérsic profile).}
    \label{fig:simulated_data}
\end{figure}

The simulated lensing data is then created by providing the mass model and source to \molet, which performs ray-tracing on a high-resolution grid, convolution with the PSF, downsampling, and adding noise to the final mock observation. We use a simulated HST PSF using \textsc{TinyTim} \citep{Krist201120YO_tinytim} based on the WFPC2 instrument with the F814W filter (we do not consider the more recent WFC3 instrument as \textsc{TinyTim} does not support it). Since ray-tracing is performed on a 10 times higher resolution grid compared to the final data, we use a simulated PSF at that resolution for more accurate surface brightness convolutions (although modelers are given a PSF at the data resolution, as is mentioned in Sect.~\ref{ssec:common_aspets}). Our settings for the noise correspond to 2200s of exposure in the chosen instrument setup. The angular size of the source cutout is set to 4 arcsec, and we scale its total flux such that it has an apparent (unlensed) AB magnitude of 23.2, which is in the range of observed source galaxies from the Sloan Lens ACS Survey (SLACS) sample \citep{Bolton2006,Newton2011}. We show in the top left panel of Fig.~\ref{fig:simulated_data} the simulated lens image, while the bottom row shows the supersampled and data-resolution PSFs (only the latter is provided to the modelers).

We note that the lensed source galaxy and the data S/N are such that a simple source model is not able to fit the data. We visualize this in the top right panel of Fig.~\ref{fig:simulated_data}, which shows normalized residuals between the data and a model based on a single Sérsic profile for the source. Such residuals are strong evidence for the necessity of more flexible source models as the ones we employ in this work (described in Sect. \ref{sec:modeling_methods}).

In Appendix~\ref{app:sec:previous_mock} we give useful details about a previous version of the mock we attempted to model, for which we detected issues related to the input source light distribution. In a nutshell, the original source did not have an accurate background subtraction and displayed sharp edges (visually unnoticeable after lensing the addition of noise), which led to biases in the lens models. For this reason, a second mock with different input parameters and source light had to be re-created (and the subsequent modeling re-done).

\section{Standardized comparison framework \label{sec:coolest}}

Our work relies on several collections of modeling methods and software packages, that we systematically apply on the same data. These codes have been developed following different conventions (e.g., angles, units, profile definitions), are written in different programming languages (e.g., Python, C++), and differ in their final modeling products. Therefore, we must ensure that we can both simulate and model strong lensing data in a consistent way, in order to mitigate problems arising from the heterogeneous collection of methods we consider.

We used the recently released strong gravitational lensing standard \coolest (for COde-independent Organized LEnsing STandard) as a framework unifying the different components of our analysis\footnote{\coolest is an open source Python package publicly available at \url{https://github.com/aymgal/COOLEST}. We used the released version 0.1.9.}. Below we briefly describe \coolest and its specific features that we used in this work, but refer the reader to \citet{Galan2023coolest} and the online documentation\footnote{\url{https://coolest.readthedocs.io}} for more details.

The foundation of \coolest is a set of conventions to serve as a reference point for modeling assumptions and codes, for example, coordinate systems, units and profile definitions. Given these conventions, any lens model---together with the data being modeled and other modeling components such as the PSF---can be concisely described in a single file following the JSON format. We refer to the latter as a \coolest template file, which we use both to describe an instance of a strong lens to be simulated and to store lens modeling results (e.g., best-fit parameters and uncertainties). This standard way of storing lens modeling information allows us to straightforwardly compare any modeling results to each other as well as to an existing groundtruth (in the case of simulated data). In practice, it requires each modeling or simulation code to have an interface with \coolest to create or update such template files\footnote{This task is made easier by using the dedicated \coolest Python interface.}. Lastly, we use the analysis features of \coolest to read the content of template files, compute key lensing quantities, and produce comparison plots. In particular, we use this interface to plot all lens models side-by-side and compare them to the groundtruth, compute morphological features of reconstructed source galaxies, and plot joint posterior distributions over the parameter space.

\section{Results from blind modeling with six methods \label{sec:results_mock_data}}

After agreeing on the properties of the simulated data (instrumental setup, S/N, type of mass model), one of the authors (M.G.) first simulated the imaging data with a first software (Sect.~\ref{sec:data_simulation}). The rest of the authors proceeded with the blind modeling of the data, whose results are presented here. The modeling workload was split between different modelers: L.V.V. (Sérsic+Shapelets models), G.V. (Adaptive+Matérn model), Q.M. (Cluster+Exp and Cluster+Exp+Lumweight models) and A.G. (Sparsity+Wavelets and Correlated Field models). In total, we used four distinct software packages and six different modeling methods. Each modeler was free to perform their analysis to the best of their judgment. Once confident that no significant improvements to the models could be made by further fine-tuning, the modelers converted their results to the \coolest format (Sect.~\ref{sec:coolest}) and submitted them to A.G. (this may have been a single or a marginalization of model instances). During the simulation and the subsequent modeling phase, there was minimum amount of information shared between the authors. In particular, the shared information between the simulating author and the modelers was restricted to the only data pixels, the corresponding noise map as well as the data-resolution PSF. Afterwards, the modelers did not share their lens modeling results before all authors agree to do so. Keeping this part of the analysis blind ensured unbiased and independent models. Unblinding took place when M.G. also submitted the true model, and the final figures presented in this section were produced. The specific steps necessary to obtain these models, as well as their estimated computation times, are detailed in Appendix~\ref{app:sec:model_details}. All models along with the simulated data products are publicly available\footnote{\url{https://github.com/aymgal/LensSourceDegeneracy_public}. Upon acceptance of the manuscript, analysis and visualization notebooks will also be released.}.

\subsection{Overall fit to the imaging data}

\begin{figure*}[!h]
    \centering
    \includegraphics[width=\linewidth]{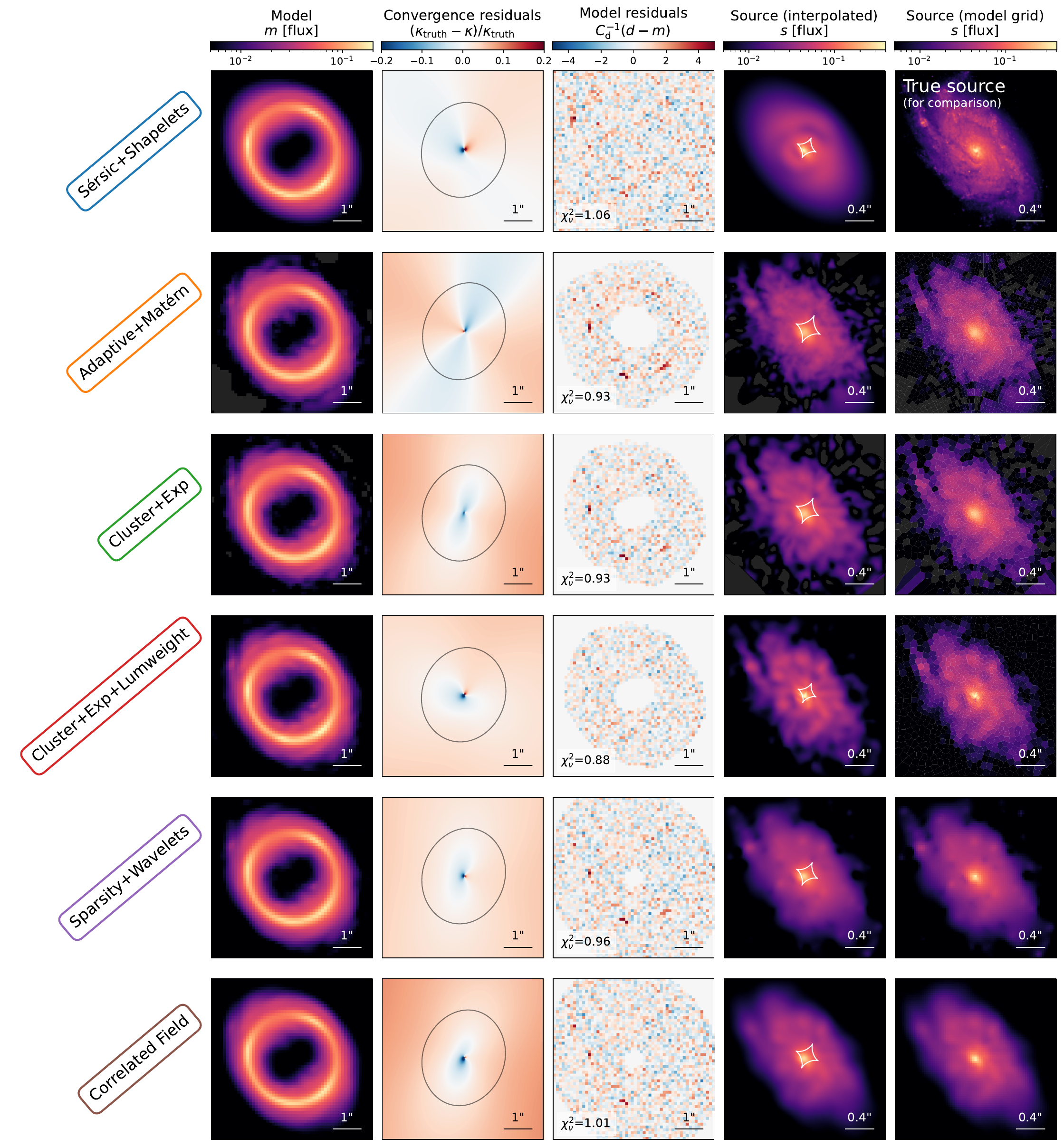}
    \caption{Comparison between all blindly submitted models of the data shown in Fig.~\ref{fig:simulated_data}. The leftmost area gives the labels associated with each modeling method (see also Table~\ref{tab:modeling_methods}), as well as their associated color used for the subsequent figures of this paper. \textit{First column}: Image model. \textit{Second column}: Relative difference between the true and modeled convergence maps, with the predicted tangential critical line shown as a black line. \textit{Third column}: Normalized model residuals, with associated reduced chi-squared $\chi_\nu^2$ indicated in the bottom left. The white areas are outside the likelihood mask chosen by the modelers and are thus excluded during modeling. \textit{Fourth column}: reconstructed source models, all interpolated onto the same (regular) high-resolution grid to ease visual comparison. The predicted tangential caustic is also indicated as a white line. \textit{Last column, first panel}: True source as in Fig.~\ref{fig:simulated_data}, to ease comparison with the models. \textit{Last column, remaining panels}: Reconstructed source models on their original discretization grid, which can be regular or irregular, when applicable (the Sérsic+Shapelets model is not defined on a grid). All panels have been generated using \coolest routines from the standardized storage of each model.}
    \label{fig:sim_data_comp_models}
\end{figure*}

We show in Fig.~\ref{fig:sim_data_comp_models} all the six models that have been blindly submitted, in direct comparison with the true model from the simulation. We compare side-to-side the image model, relative error on the convergence, model normalized residuals (in unit of the noise), and the reconstructed sources. The reconstructed sources should be directly compared to the top right panel of Fig.~\ref{fig:simulated_data}. The residuals show clear improvements over the residuals shown in the bottom right panel of Fig.~\ref{fig:simulated_data} with a too simplistic source model. We also quote the reduced chi-square value $\chi_\nu^2$ computed within the likelihood mask chosen by the modeler, in order to quantitatively compare the quality of fits. The largest $\chi_\nu^2$ value is 1.06 while the lowest value is 0.88, and four models have a $\chi_\nu^2$ below unity which indicate slight overfitting. In several models, residuals at the $\sim 3\sigma$ level remain where the arcs are the brightest (multiple images of the bright, cuspy central region of the source). We note that overall, all models fit the imaging data to very close to noise level.

While the Sérsic+Shapelets model captures less small-scale structures in the source compared to other models, the resulting fit in the image plane still achieves noise-level residuals (except maybe for a few pixels in the outskirts of the lensed source). Finer structures in the source like spiral arms and star forming regions are overall better modeled by semi-linear inversion methods, than by the wavelets and the correlated field.

The second column of Fig.~\ref{fig:sim_data_comp_models} shows the relative error in the lens convergence, throughout the field-of-view. Offsets in the lens centroid and in the ellipticity are visible, although strongest errors remain at the very center of the lens. At the position of the multiple images (roughly traced by the critical lines), the relative error in convergence remains below 5\%.

\begin{figure}
    \centering
    \includegraphics[width=\linewidth]{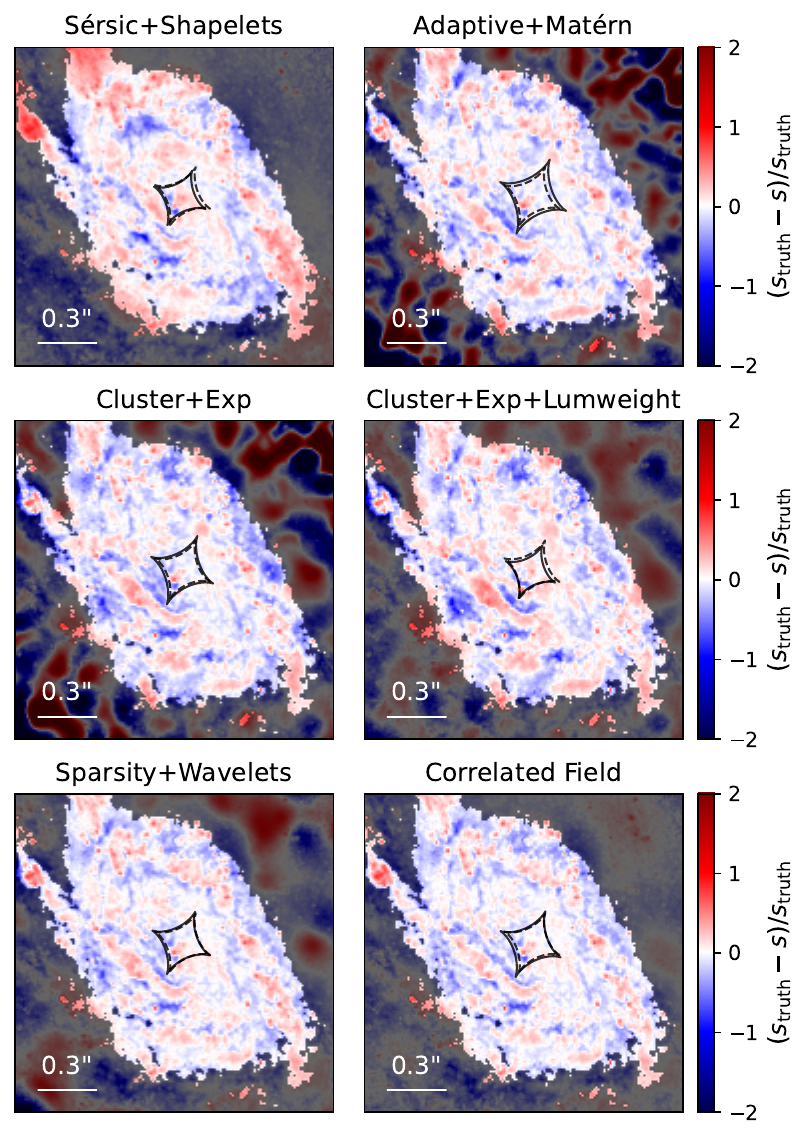}
    \caption{Source plane residuals between the true source and the models shown in the fourth row of Fig.~\ref{fig:sim_data_comp_models}. Away from the center, the true source intensity is close to zero and models are less accurate, thus these areas are darkened for better visualization. Caustics from the best-fit (solid lines) and true (dashed lines) lens models are shown as well. Within the caustics, all models overall recover the source structure.}
    \label{fig:sim_data_comp_src_diff}
\end{figure}

\subsection{Recovery of source properties \label{ssec:results_source}}

To visualize better which source features are captured by the models, we show in Fig.~\ref{fig:sim_data_comp_src_diff}, maps of source plane residuals computed as the difference between the true and reconstructed sources. Far from the optical axis, models are not well constrained by the data and significantly deviates from the true light distribution; thus, we darken these areas for better visualization. Pixelated models defined on irregular grids capture similarly well most of the spiral features as well as star forming region located in the left spiral arm. The center of the source galaxy seems to be retrieved equally well by all models, despite the persistent image plane residuals (see Fig.~\ref{fig:sim_data_comp_models}).

\begin{figure*}
    \centering
    \includegraphics[width=0.455\linewidth]{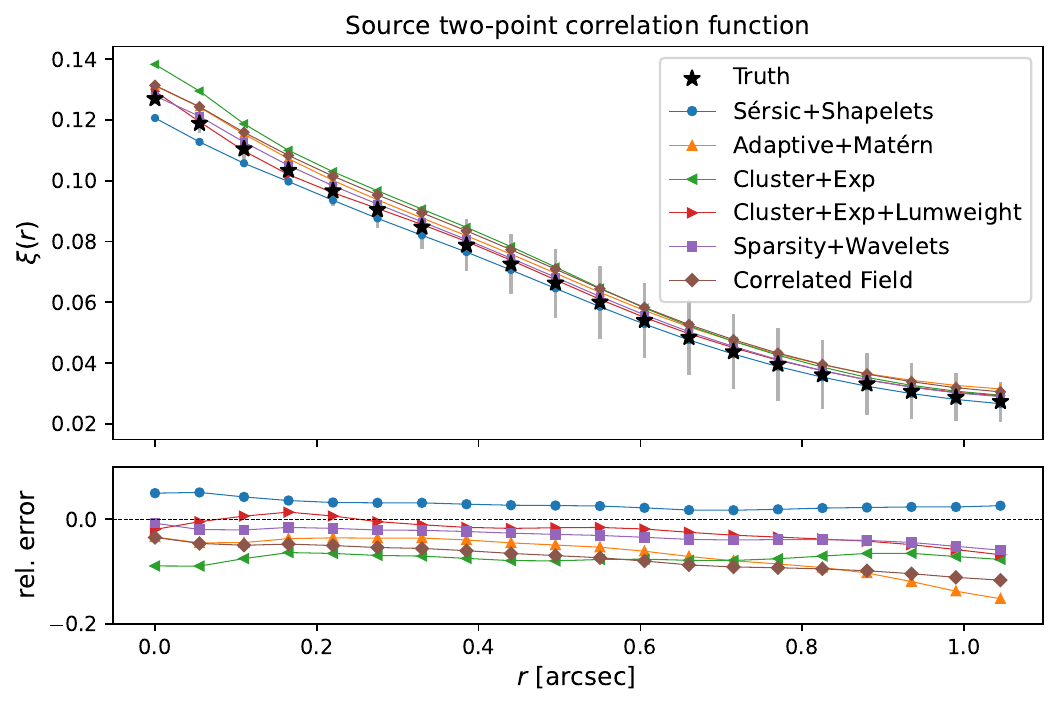}
    \includegraphics[width=0.485\linewidth]{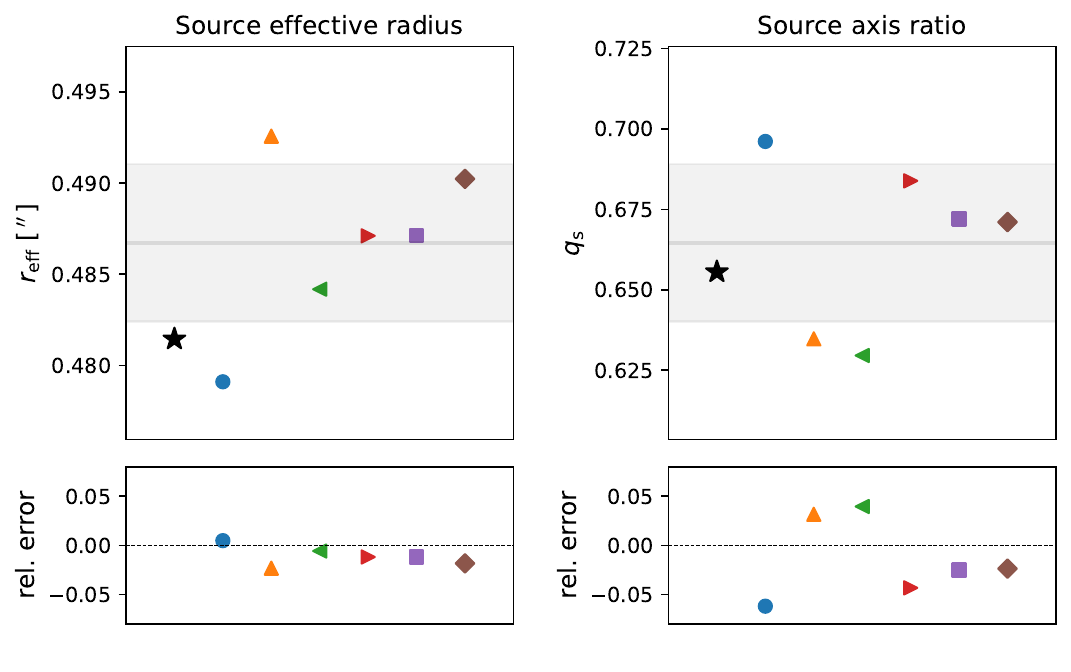}
    
    \caption{Recovery of source properties among the six blind lens models shown in Fig.~\ref{fig:sim_data_comp_models} (the legend in the right panel is also valid of other panels). \textit{Left panels}: Azimuthally averaged two-point correlation functions, $\xi(r)$, of the recovered sources, compared to the truth, which measures the correlation between pixels separated by a given distance, $r$ (in arcsec). \textit{Middle panels}: Source effective radius $r_{\rm eff}$, computed within a circular aperture. \textit{Right panels}: Source axis ratio, $q_{\rm s}$, estimated from the central moments of the source model. The bottom panel in each column indicates, the relative error with respect to to the true quantity, computed as $(\rm{truth} - \rm{model})/\rm{truth}$. For $r_{\rm eff}$ and $q_{\rm s}$, the shaded gray region indicates the mean and $1\sigma$ scatter among the models.}
    \label{fig:sim_data_comp_src_props}
\end{figure*}

We show in Fig.~\ref{fig:sim_data_comp_src_props} the recovery of several properties of the source galaxy: the two-point correlation function, the effective radius and the axis ratio. Bottom parts of each panel show the relative error computed as $(\rm{truth} - \rm{model})/\rm{truth}$ (i.e., negative values are over-estimates). We measure these properties within a square field of view of size $2\farcs2$, after projecting (using bi-cubic interpolation) each source model on a regular grid with 10 times higher resolution than the data. As is shown in Fig.~\ref{fig:sim_data_comp_models}, such a field of view contains the entire flux of the source galaxy.

The left panel of Fig.~\ref{fig:sim_data_comp_src_props} shows the two-point correlation function $\xi(r)$, which gives the azimuthally averaged correlation between the source intensity at two positions in source plane, as a function of their angular separation. All reconstructed sources exhibit two-point correlations close to the one of the input galaxy. Over all the models, the maximum error remains small, except for some models for which it exceeds $15\%$ error on two-point correlations on arcsecond scales. We note that the Sérsic+Shapelets model under-estimates two-point correlations at all scales, while other models over-estimate these correlations. The Cluster+Exp+Lumweight reaches minimal error on the smallest scales ($\lesssim0\farcs4$), which shows that the more detailed reconstruction obtained with this model, visible in Fig.~\ref{fig:sim_data_comp_models}, is accurate over these small spatial scales.

We investigate the recovery of the size of the source galaxy through its effective radius, $r_{\rm eff}$, which we define as the radius that encloses half of the total light within a circular aperture of radius $2\farcs2$. The middle panel of Fig.~\ref{fig:sim_data_comp_src_props} shows $r_{\rm eff}$ and its relative error with respect to the true value for all source models. The effective radius is well recovered by all models with a maximum error of $2.5\%$. In addition we observe a tendency to over-estimate the effective radius, as only the Cluster+Exp model slightly under-estimates it, which is also the model with the lowest error. However, a better quantification of these errors should involve posterior distributions over the source models, which we do explore in this work. A first order uncertainty quantification can be obtained through the scatter among the different models, shown as the shaded gray area in Fig.~\ref{fig:sim_data_comp_src_props}. Given this scatter, the average source half-light radius remains lies close to $1\sigma$ from the true value.

Over the different source models considered here, none explicitly parametrizes the ellipticity of the source galaxy, in particular its axis ratio. Therefore, we use central moments of source model images (projected onto the same coordinates), in order to empirically measure an axis ratio $q_{\rm s}$. More specifically, we compute the second order central moments of the source image, and use its eigenvalues to estimate the major and minor axes, from which we obtain the axis ratio $q_{\rm s}$. The rightmost panel of Fig.~\ref{fig:sim_data_comp_src_props} shows the resulting values, along with the true value measured on the true source image. The relative error is overall larger than for the effective radius although it remains below $6\%$. Taking the mean over the ensemble of modeling methods lies very close to the true value, within the $1\sigma$ scatter among the models.

\begin{figure}
    \centering
    \includegraphics[width=\linewidth]{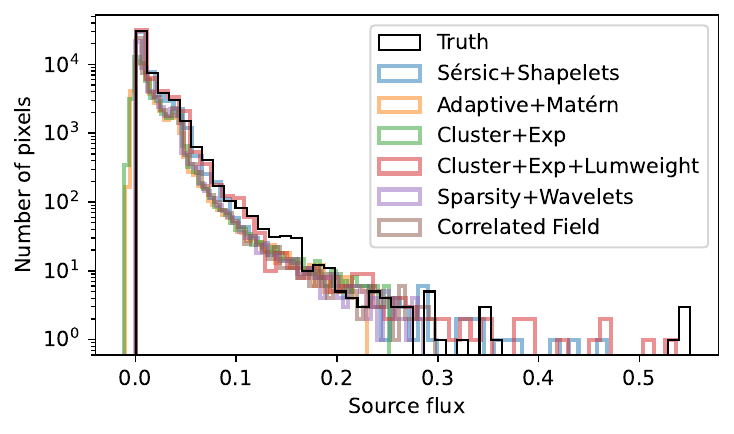}
    \caption{Histogram of pixel intensities for each reconstructed source shown in Fig.~\ref{fig:sim_data_comp_models}. Some models allow for slightly negative intensities, while some can capture very high intensity and compact source features.}
    \label{fig:sim_data_comp_src_hist}
\end{figure}

Figure~\ref{fig:sim_data_comp_src_hist} shows the histogram of source pixel intensities for each model, compared to the true intensities (after interpolation, i.e., rightmost column of Fig.~\ref{fig:sim_data_comp_models}). Interestingly, we clearly see that the Sérsic+Shapelets and Cluster+Exp+Lumweight models reach higher intensity values. This is expected for the former (Sérsic+Shapelets) as the Sérsic profile diverges (with a best-fit Sérsic index $\approx1.6$) in its center and thus can predict large flux values. The luminosity-weighted regularization of the latter (Cluster+Exp+Lumweight) is behaving similarly by decreasing the regularization strength in regions with high observed flux. Consequently, these two models are best at capturing the three most magnified images of the source (see third column of Fig.~\ref{fig:sim_data_comp_models}). We also note from Fig.~\ref{fig:sim_data_comp_src_hist} that some of the models exhibit slightly negative values, as those are not penalized in their underlying prior, although these value are not statistically significant when compared to the noise level, and located mainly on the outskirts of the reconstructed source.

\subsection{Recovery of lens properties \label{ssec:results_lens}}

\begin{figure*}
    \centering
    \includegraphics[width=\linewidth]{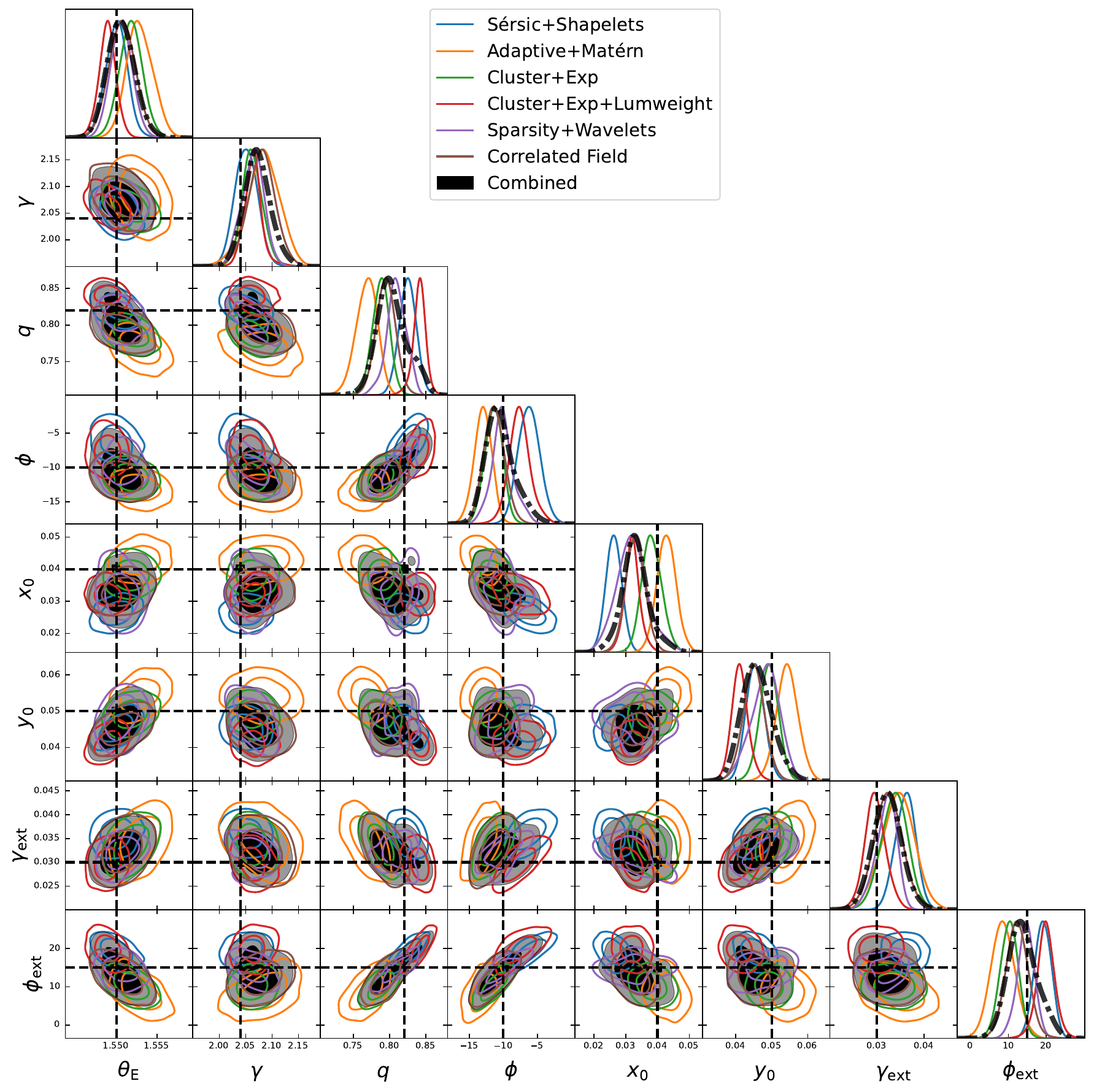}
    \caption{Posterior distributions of the lens potential parameters, inferred blindly from the simulated data shown in Fig.~\ref{fig:simulated_data}, with the true values indicated by the dashed lines. The parameters from left to right are: the Einstein radius, mass density logarithmic slope, axis ratio, position angle, lens center coordinates, external shear strength, and external shear orientation. Note that we use the \coolest definitions of these parameters.}
    \label{fig:sim_data_comp_lens_full_corner}
\end{figure*}

We investigate the different constraints on the mass distribution of our simulated strong lens system obtained from the different modeling methods. Already on Fig.~\ref{fig:sim_data_comp_src_diff}, showing the predicted tangential caustics, we can visualize slight differences among the models. The predicted caustics have different sizes and positions, overall slightly larger than the true caustics. These differences correlate with the corresponding biases we discuss below. In particular, larger predicted density slopes are responsible for increasing the caustic size, and lens ellipticity and position offsets both impact the position and orientation of the astroid.

To quantitatively compare the constraints on lens potential parameters among the six models, we show in Fig.~\ref{fig:sim_data_comp_lens_full_corner} the joint posterior distributions for all mass model parameters, as well as true values from the data. Overall we find that the posterior distributions are within $\sim3\sigma$ from the truth. The parameter with the smallest scatter relative to the posterior width is the logarithmic density slope $\gamma$. While all models are slightly biased towards values larger than the true value by approximately $2\%$, they are all compatible with it at $\lesssim1\sigma$. Such a low scatter in the slope may be perhaps surprising, as it has been shown that the density slope may differ significantly between models \citep[e.g.,][]{Etherington2022,Tan2024}. However, we are in a regime where both the data and all models are parametrized by a single power-law density profile, hence the data is by construction a realization of the true model (modulo our inexact knowledge of the PSF and different numerics settings). Besides the density slope, the Einstein radius, $\theta_{\rm E}$, also has low scatter among the models, which is expected as it is the primary quantity constrained by strong lensing observables. Models whose median values are further away from the truth also tend to have broader posteriors (larger uncertainties), which contribute to reduce the systematic bias (see also Fig.~\ref{fig:sim_data_comp_lens_uncert} below). The remaining mass model parameters all show visible scatter around the true values, while no systematic shift can be associated with a specific modeling method or model parameter.

\begin{figure}
    \centering
    \includegraphics[width=\linewidth]{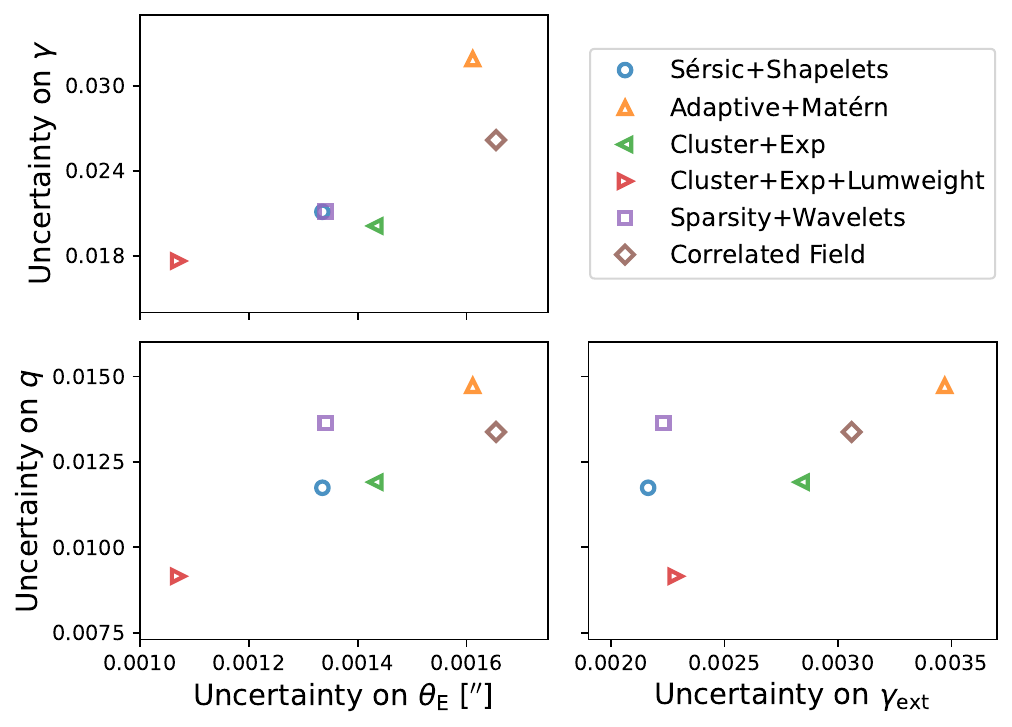}
    \caption{Uncertainty on a subset of key lens potential parameters, defined as the standard deviation of their marginalized posterior distribution (shown in Fig.~\ref{fig:sim_data_comp_lens_full_corner}): the Einstein radius, $\theta_{\rm E}$, the mass density slope, $\gamma$, the axis ratio $q$, and the external shear strength $\gamma_{\rm ext}$. The Einstein radius and the mass density slope are central quantities to many analyses, while the lens axis ratio and external shear are known to share degeneracies.}
    \label{fig:sim_data_comp_lens_uncert}
\end{figure}

From the full joint distributions over lens potential parameters, we investigate further the difference in uncertainties between the models. In Fig.~\ref{fig:sim_data_comp_lens_uncert} we plot posterior standard deviations for four parameters, $\theta_{\rm E}$, $\gamma$, $q$, and $\gamma_{\rm ext}$. As is expected from imaging lensing data, the uncertainty on $\theta_{\rm E}$ is the smallest, with a relative precision on the order of $0.1\%$. For comparison, the relative precision on the mass density slope $\gamma$ is around $1.2\%$, and around to $1.5\%$ for the lens axis ratio $q$ (the relative error for $\gamma_{\rm ext}$ is inconclusive since it is close to zero). We notice that models with largest dynamic ranges (see Sect.~\ref{ssec:results_source}) in their reconstructed source (Cluster+Exp+Lumweight) have smaller uncertainties on lens potential parameters. This trend is particularly clear for $\theta_{\rm E}$, $\gamma$, $q$ (first column in Fig.~\ref{fig:sim_data_comp_lens_uncert}). Over the parameters shown in Fig.~\ref{fig:sim_data_comp_lens_uncert}, the difference in uncertainties between the models remains relatively small, and amounts to a factor of approximately 1.6 between the least and most precise models.

\begin{table*}
    \caption{Posterior statistics of source and lens properties inferred from the six lens models shown in Fig.~\ref{fig:simulated_data}.}
    \label{tab:scatter_models}
    \renewcommand{\arraystretch}{1.8}
    \centering
    \small
    \begin{tabular}{cccccccccccc}
        \hline\hline
         Quantity & $r_{\rm eff}$ [$''$] \tablefootmark{$\dag$} & $q_{\rm s}$ \tablefootmark{$\dag$} & $m_{\rm s}$ [mag] & $\theta_{\rm E}$ [$''$] & $\gamma$ & $q_{\rm m}$ & $\phi_{\rm m}$ & $x_{\rm 0}$ [$''$] & $y_{\rm 0}$ [$''$] & $\gamma_{\rm ext}$ & $\phi_{\rm ext}$ [$^\circ$] \\
        \hline
        Truth                & $0.481$ & $0.66$ & $23.2$ & $1.550$ & $2.040$ & $0.82$ & $-10.0$ & $0.040$ & $0.050$ & $0.030$ & $15.0$ \\
        $\mu_{\rm models}$    & $0.487$ & $0.67$ & $23.1$ & $1.551$ & $2.066$ & $0.80$ & $-9.9$  & $0.034$ & $0.047$ & $0.033$ & $13.9$ \\
        $\sigma_{\rm models}$ & $0.004$ & $0.02$ & $0.04$ & $0.001$ & $0.009$ & $0.03$ & $2.3$   & $0.005$ & $0.004$ & $0.002$ & $4.6$  \\

        Max. bias\tablefootmark{$\dag\dag$} & \textcolor{gray}{$\times$} & \textcolor{gray}{$\times$} & \textcolor{gray}{$\times$} & $1.7\sigma$ & $1.6\sigma$ & $-3.3\sigma$ & $2.3\sigma$ & $-5.5\sigma$ & $-3.7\sigma$ & $2.7\sigma$ & $2.2\sigma$ \\

        \hline
        
        Combined              & \textcolor{gray}{$\times$} & \textcolor{gray}{$\times$} & \textcolor{gray}{$\times$} & $1.55_{-0.002}^{+0.002}$ & $2.07_{-0.02}^{+0.03}$ & $0.80_{-0.03}^{+0.03}$ & $-10.5_{-2.6}^{+1.4}$ & $0.033_{-0.004}^{+0.004}$ & $0.046_{-0.005}^{+0.004}$ & $0.032_{-0.003}^{+0.003}$ & $13.8_{-4.5}^{+3.3}$ \\

        Bias\tablefootmark{$\dag\dag$} & \textcolor{gray}{$\times$} & \textcolor{gray}{$\times$} & \textcolor{gray}{$\times$} & $0.3\sigma$ & $1.2\sigma$ & $-0.6\sigma$ & $-0.2\sigma$ & $-1.6\sigma$ & $-0.9\sigma$ & $0.7\sigma$ & $-0.3\sigma$ \\
        
        \hline
    \end{tabular}
    \tablefoot{
    In the second and third row $\mu_{\rm model}$ and $\sigma_{\rm model}$ show the average and standard deviation of the posterior means of all models, respectively. The fourth row gives the maximal bias among all models for each lens potential parameter. The two last rows correspond to the combined posterior distribution shown in black in Fig.~\ref{fig:sim_data_comp_lens_full_corner}.\\
    \tablefoottext{$\dag$}{The source properties correspond to best-fit models, whereas lens properties relate to the corresponding marginalized posterior distributions shown in Fig.~\ref{fig:sim_data_comp_lens_full_corner} (individual and combined models). Note that we do not provide combined estimates for the source properties as we do not have formal posterior distributions not uncertainty estimations for all source models.}\\
    \tablefoottext{$\dag\dag$}{The maximum bias among all models is based on the difference between the mean posterior value and the true value, in units of the posterior standard deviation (negative and positive biases correspond to under- or over-estimates, respectively). The last row of the table shows the corresponding bias value from the combined posterior distribution.}
    }
\end{table*}

Ensemble models---namely, the combination of the posteriors of multiple models---can help improve modeling accuracy by correcting for the observed systematic biases of individual models. Depending on the assumptions about the original individual models, in particular regarding their statistical independence, there exist different approaches to combine their posteriors together. Here we follow a conservative approach and simply combine individual posteriors with equal weights. We show with dash-dotted black lines and contours in Fig.~\ref{fig:sim_data_comp_lens_full_corner} the resulting combined posteriors. We find that these combined distributions are all within $1\sigma$ from the true values, except for the lens center along the $x$ direction which is at $1.7\sigma$. The marginalized statistics of the combined posterior are reported in the last row of Table~\ref{tab:scatter_models} and compared to the simple average and standard deviation among individual models.

We quantify the improvement in systematic bias between individual models and the combined model. In the fourth row of Table~\ref{tab:scatter_models}, we list the largest bias (in units of standard deviation) that arises among the six lens models, for each lens potential parameter. As already seen in Fig.~\ref{fig:sim_data_comp_lens_full_corner}, the most biased parameters appear to be the coordinates of the center of the power-law profile, $(x_0,y_0)$, while the least biased is the density slope $\gamma$. As a comparison, the last row of Table~\ref{tab:scatter_models} lists the corresponding bias values of the combined posterior, showing as expected a substantial decrease for all mass model parameters. We discuss further these results in Sect.~\ref{sec:discussion}.

\subsection{Correlations between the recovery of lens and source properties \label{ssec:lens_source_correl}}

Having in hand multiple lens models of the same data, we have the opportunity to explore how the accuracy of inferred lens and source properties correlate among the models. In particular, it is interesting to understand if certain biases observed in lens potential model parameters have an origin in biases in the reconstructed source light distribution, and vice versa. If such correlations exist, they could be used to design better parametrizations to jointly model the lens and source components that specifically break these degeneracies. Such degeneracies may also be broken using non-lensing observations to further improve the accuracy of inferred lens and source properties (e.g., stellar kinematics of the source galaxy to place complementary constraints on its morphology). We emphasize that such correlations may be system- and data-dependent, which would warrant additional analyses complementary to ours.

We show in Figs.~\ref{app:fig:correl_r_eff} to \ref{app:fig:correl_mag} a series of scatter plots that correlate the relative error on lens potential parameters with the relative error on a given source property. We compute uncertainties on lens potential parameters from their posterior standard deviation. As we do not have such posterior distributions for all source models, we assume a fiducial uncertainty based on the Correlated Field model, for which we have posterior samples (see Sect.~\ref{ssec:general_fields}).

To quantify possible correlations, we compute the biweight mid-correlation coefficient $r$, indicated on each panel in Figs.~\ref{app:fig:correl_r_eff} to \ref{app:fig:correl_mag}. The uncertainty on $r$ is estimated by drawing 1\,000 random samples from bivariate uncorrelated Gaussian distributions centered on each data point. Based on the biweight mid-correlation coefficients, the largest (anti) correlation arises between the $x$-coordinate of the lens centroid $x_0$ and the axis ratio of the source $q_{\rm s}$, with $r=-1.0\pm0.2$. On the other hand, the strongest absence of correlation is seen between the external shear strength $\gamma_{\rm ext}$, and the total source magnitude $m_{\rm s}$, with $r=0.1\pm0.2$. We discuss and interpret the observed correlations in Sect.~\ref{sec:discussion}.

\section{Investigating sources of systematics \label{sec:systematics_tests}}

While a thorough investigation of all possible sources of systematics is beyond the scope of this study, we nevertheless attempt to assess the impact of some key modeling assumptions and data properties.
The results of this effort will be useful in guiding future in-depth investigations.
Specifically, we explore the role of the knowledge of the lens position, the presence of small-scale high-contrast regions (cusps, or point-like features) in the light profile of the source galaxy, and imperfect knowledge of the PSF.

\subsection{Intrinsic source morphology}
The spiral galaxy light profile that was used as the lensed source in producing the mock data presented in Sect. \ref{sec:data_simulation}, has a prominent bright spot its center. The best-fit Sérsic index from the Sérsic+Shapelets model is approximately $1.6$, indicative of a cuspy radial profile.
This feature, which we shall refer to as the source cusp in the following paragraphs, consists of only a handful of pixels (<20) that contain a significant amount of light ($\sim$5 \%).
This region is clearly hard to model, as is shown by the reconstructions in the last two columns of Fig. \ref{fig:sim_data_comp_models}, where only two models are able to capture it (S{\'e}rsic+Shapelets and Cluster+Exp+Lumweight).
The remaining models fail to do so and leave behind characteristic residual flux at the data pixels where this compact region is multiply imaged.

Driven by this observation, we argue that this cusp is closer to a point-like flux component than to an extended source.
Some algorithms, like the plain semi-linear inversion (even on an adaptive grid), have been explicitly designed to model the latter and are known to have a poor performance with the former.
This can be understood in terms of the regularization, which tries to impose smoothness on the source and inevitably suppresses such cusps.
In Fig. \ref{fig:sim_data_comp_src_hist}, we see that only a few source pixels with the highest flux ($>0.3$), corresponding to the central cusp, are considerably extending the dynamic range of the source light profile.
The two aforementioned models that successfully model the cusp have a similar dynamic range, while the rest of the models fall short, barely reaching a flux of 0.3.

\begin{figure}
    \centering
    \includegraphics[width=0.9\linewidth]{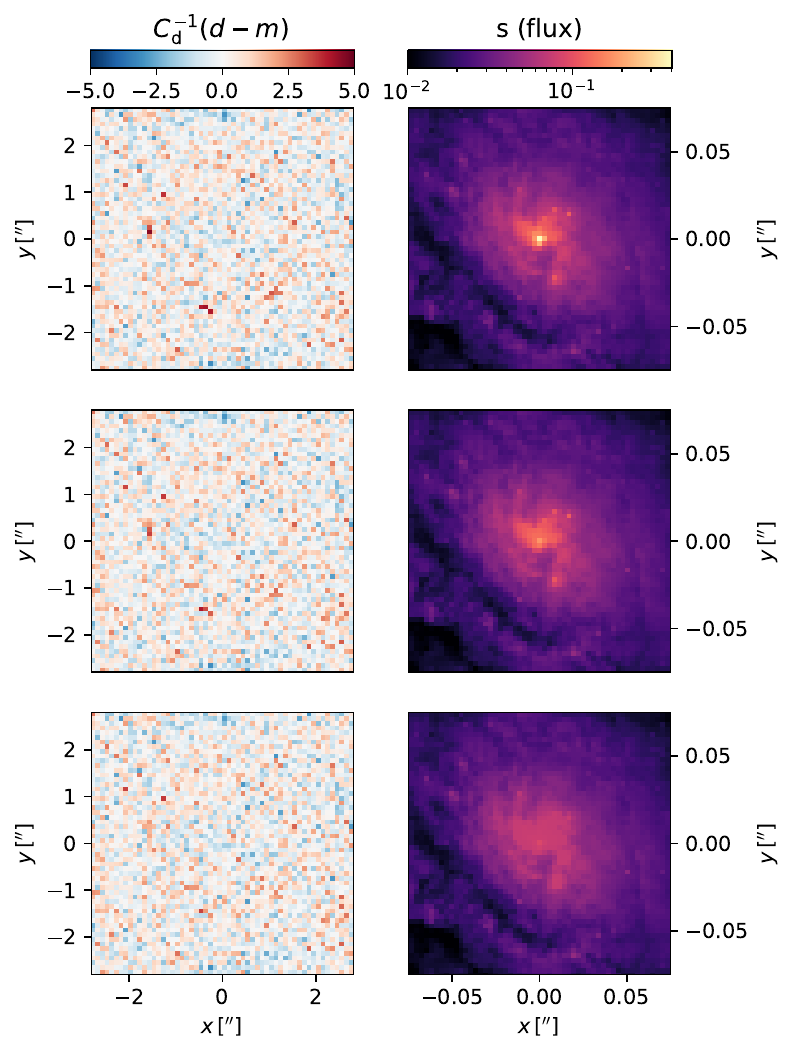}
    \caption{Model residuals from the Adaptive+Matérn model and central region of the true source (top row, as in Fig.~\ref{fig:sim_data_comp_models}). Suppressing the flux in the central most bright pixels ($>0.3$ in Fig.~\ref{fig:sim_data_comp_src_hist}) by 50 and 95 \% (middle and bottom rows), creating new mocks, and modeling them with the same setup, leads to improved residuals. The plain semi-linear inversion technique using just a regularization of the form given in Eq.~\ref{eq:quadratic_reg_general} cannot adequately capture point-like, cuspy features in the source light profile. More advanced schemes, like a luminosity-weighted regularization scheme (see Sect. \ref{ssec:general_semlinear_inv}), perform better in this case.}
    \label{fig:cusp_residuals}
\end{figure}

\begin{figure}
    \centering
    \includegraphics[width=\linewidth]{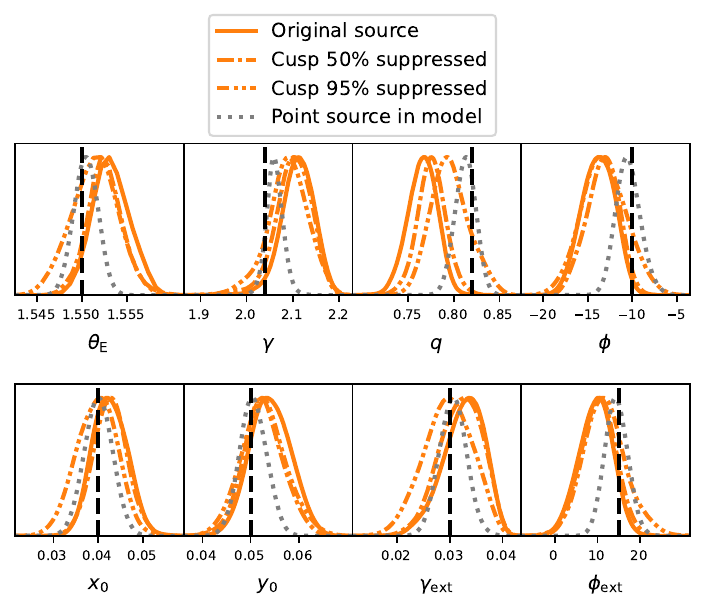}
    \caption{Posterior distributions of the lens potential parameters for the three models shown in Fig.~\ref{fig:cusp_residuals}. The fourth model (bottom in the legend) corresponds to a model of the original mock after including a point source component in the source model. When the cuspy central region of the source is suppressed from the data, or if a point source feature is added in the source model, the resulting distributions are less biased and shift closer to the true values.}
    \label{fig:cusp_corner}
\end{figure}

Here, we examine in more details how the performance of the model with the lowest brightness range (smoothest), Adaptive+Matérn, is affected by the prominence of the cusp.
We select the central bright region with flux $>0.3$ and reduce the flux\footnote{The noise map that is used as the covariance matrix in Eq.~\ref{eq:data_likelihood} is also changed accordingly.} in each pixel by 50 and 95 \%, creating two new mocks that we then model with exactly the same setup as the model shown in Fig. \ref{fig:sim_data_comp_models}.
The resulting (true) source and model residuals are shown in Fig. \ref{fig:cusp_residuals}, while Fig. \ref{fig:cusp_corner} shows the posterior distributions of the lens potential parameters.
It can be seen that the more we suppress the cusp the better the model; that is, we get less residual flux and less biased parameters.

We conclude that a plain semi-linear inversion approach is much better suited for modeling smoother sources, without cuspy, point-like features in their light profile.
A regularization scheme imposed just by Eq. \ref{eq:quadratic_reg_general} leads to reconstructions that are too smooth, and additional constraints, like the luminosity-weighted scheme presented in Sect. \ref{ssec:general_semlinear_inv}, give better results.
The cuspy nature of the source can thus explain, at least partially, the systematic errors in lens potential parameters, in particular for the case of the Adaptive+Matérn model.

\subsection{Supersampling and inexact PSF model} \label{sec:supersampling}

\begin{figure}
    \centering
    \includegraphics[width=\linewidth]{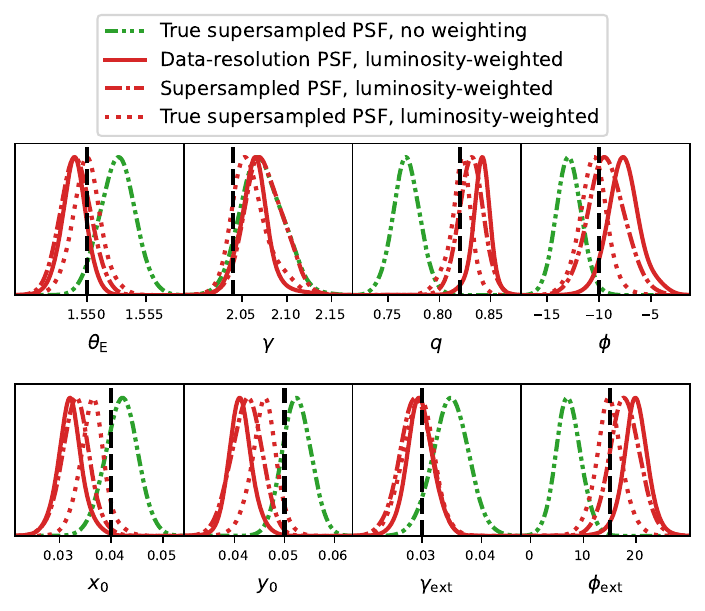}
    \caption{Posterior distributions of lens potential parameters obtained with models exploring the role of the PSF. The dash-dotted green line distributions correspond to the Cluster+Exp model (i.e., similar to the green model in Fig.~\ref{fig:sim_data_comp_lens_full_corner}) using the true supersampled PSF used for simulating the data (top left panel of Fig.~\ref{fig:psf_kernels}). The solid red line distributions are showing, for reference, the blindly submitted model Cluster+Exp+Lumweight (i.e., the same model as in Fig.~\ref{fig:sim_data_comp_lens_full_corner}). The dash-dotted and dotted red distributions, also obtained with the Cluster+Exp+Lumweight model, use a supersampled (interpolated) version of the data-resolution PSF (bottom left panel in Fig.~\ref{fig:psf_kernels}) or the true supersampled PSF, respectively. The main result of this comparison is that biases in lens model parameters are most reduced only with a combination of a more accurate PSF and a model that can capture magnified cuspy features in the source.}
    \label{fig:psf_effect}
\end{figure}

\begin{figure}
    \centering
    \includegraphics[width=\linewidth]{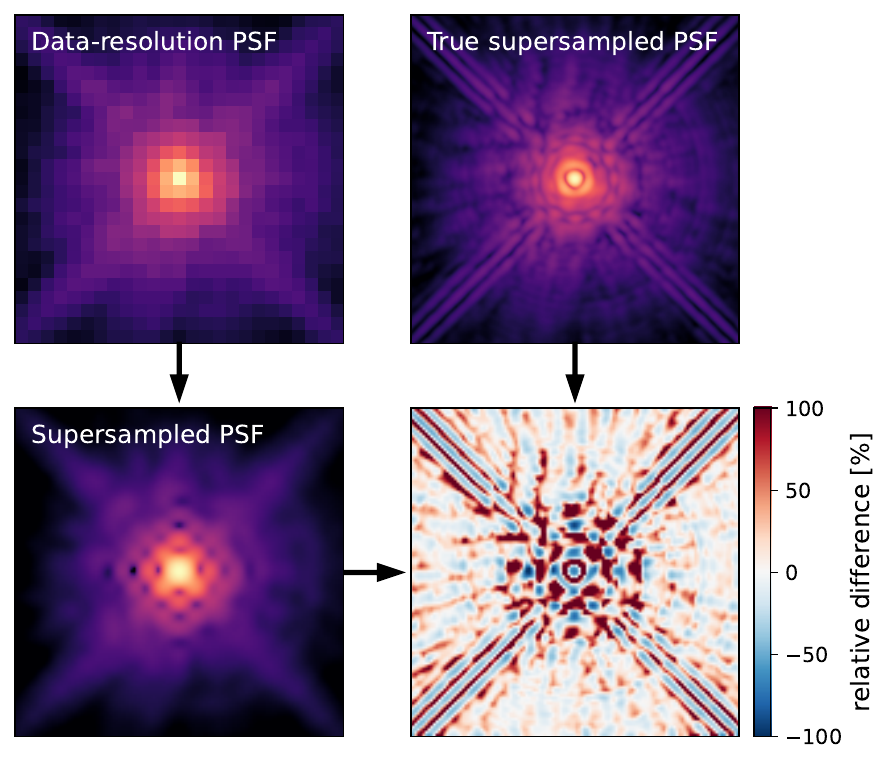}
    \caption{Comparison between the different PSF kernels used in this work (each panel shows a zoom on the central $2\farcs5$ of the PSF kernel). The top left panel shows the original true PSF used for simulating the data, while the top right panel shows the PSF given to the modelers (same as in Fig.~\ref{fig:simulated_data}). The bottom left panel shows the bicubic-interpolated (5 times supersampled) PSF used in the model shown with dotted lines in Fig.~\ref{fig:psf_effect}, and the bottom right panel shows the relative difference with the true PSF. Except for the bottom right panel, all color scales are arbitrary and chosen to help the visual comparison between the PSF kernels.}
    \label{fig:psf_kernels}
\end{figure}

We have seen in the previous section that a cuspy source can lead to biases in the lens potential parameters even in models that fit the central cusp well, in particular the Cluster+Exp+Lumweight model which achieved noise-level residuals. We now investigate whether using a supersampled PSF can further eliminate this bias. We implement PSF supersampling in two different ways: first, by interpolating in the pixel-level PSF (which is often implemented in practice when point sources are present in a lensed source); and second, by using the original supersampled PSF that was used to produce the mock data. Although the original PSF corresponds to a supersampling factor of 10 (i.e., each pixel is split into $10\times 10$ subpixels), this would be much too computationally expensive to employ directly. We therefore choose a supersampling factor of $f=5$ and downsample the original PSF with $f=10$ accordingly; this is what we shall refer to as the ``true supersampled PSF.'' In the first model, we use bicubic interpolation in the pixel-level PSF to generate our interpolated $f=5$ supersampled PSF.

We first check whether PSF supersampling can eliminate bias in models without a luminosity-weighted regularization, by applying the above supersampling procedure with the ``true supersampled PSF'' to the Cluster+Exp model. The resulting posteriors are plotted as the dot-dashed green curves in Fig.~\ref{fig:psf_effect}. In this case we find a similar bias in the lens parameters as in the case where no supersampling is performed, which is perhaps explained by the fact that the best-fit model produces similar residuals in the regions where the lensed images are brightest. This is due to the fact that the reconstructed source is not significantly better resolved without the luminosity-weighted regularization prior; the regularization strength is too high to allow for a cuspy source in the central high-intensity region of the source.

Next, we apply PSF supersampling to the corresponding model with luminosity-weighted regularization (Cluster+Exp+Lumweight). The results are shown as the red curves in Fig.~\ref{fig:psf_effect}, with the solid red curve showing the original luminosity-weighted model (i.e., same as Fig.~\ref{fig:sim_data_comp_lens_full_corner})  and the dashed and dot-dashed curves showing the models with supersampled PSF and ``true supersampled PSF'', respectively. Note that both supersampled models have significantly reduced bias compared to the original luminosity-weighted model that did not use PSF supersampling. Some bias is still present in the lens model parameters for the interpolated PSF model, particularly in the slope $\gamma$ and $\phi_{ext}$ parameters, whereas in the ``true PSF'' model, bias is largely eliminated in all lens parameters except for the center coordinates. In addition, the parameter uncertainties are significantly reduced when the true PSF is used. The essential difference can be seen by comparing the two supersampled PSF's directly in Fig.~\ref{fig:psf_kernels}, where the interpolated PSF is quite poorly resolved. We conclude that for sufficiently cuspy sources such as this one, supersampling can significantly reduce bias in the lens model parameters, but may not entirely eliminate bias in the lens parameters if one generates a supersampled PSF by interpolating in the observed pixel-level PSF.

It is interesting that without a luminosity-weighted source prior, the lens center coordinates are \emph{more} accurately recovered (regardless of whether supersampling is used), despite all the other lens parameters being significantly biased. The Sérsic+Shapelet model produced a similar bias in the lens center coordinates as the luminosity-weighted models did, which is noteworthy since these are the only models that were able to reproduce the central cusp in the source well. While it is unclear exactly why this is the case, in real applications this may be ameliorated by the fact that the foreground lens light can furnish a prior in the lens center coordinates. Aside from this caveat, the bias in these parameters is at least somewhat reduced by supersampling with the true rather than interpolated PSF. We conclude that when fitting cuspy lensed sources, PSF supersampling can significantly reduce parameter biases only if it is accompanied by a regularization scheme (e.g., luminosity-weighted) that allows the source pixels to have steeper variations in the central bright region of the source galaxy where PSF supersampling is of the greatest benefit.

\section{Discussion \label{sec:discussion}}

\subsection{Quantifying model complexity and its impact on posterior uncertainties \label{ssec:model_complexity}}

In lens modeling there is inherently a trade-off between model complexity and tractability of the final inference. On the one hand, a more complex model---namely with more model parameters, or degrees of freedom---is likely to provide a better fit to the data (commonly measured as a smaller $\chi^2$ value) with the risk of over-fitting. On the other hand, a simpler model is usually faster to optimize and often leads to a more robust inference (lower risk of local minima and multi-modal posteriors) but may not fit the data well. Moreover, models with different complexity can still fit the same data seemingly equally well; in such a case, one typically invokes the principle of Occam's razor, and prefer the one being the least complex a priori (which, when feasible, takes the form of the Bayesian evidence or other proxys such as the Bayesian information criterion).

In our work, as is described in Sect.~\ref{sec:modeling_methods}, one of the main differences between the modeling methods we consider is the way the source galaxy is reconstructed. In the past, several works focused on consistently comparing a set of lens models with different source reconstruction techniques developed under the same general formalism \citep[i.e., the semi-linear inversion formalism, see Sect.~\ref{ssec:general_semlinear_inv}, and][]{WarrenDye2003}, but differing in their regularization terms \citep[e.g.,][]{Suyu2006,TagoreKeeton2014}. Unfortunately, in the present work, there is no clear way to quantitatively and unambiguously rank the complexity of the source models considered here, given their fundamental differences in terms of mathematical formalism and underlying assumptions about the morphology of galaxies. One possibility would be to count their number of degrees of freedom, but this quantity is not readily accessible for all models. In regularized pixelated source models, the effective number of degrees of freedom is lower than them total number of source pixels, as regularization correlates source pixels over different spatial scales \citep[e.g.,][]{Suyu2006,Nightingale2015}. As a concrete example, the regularization based on sparsity and wavelets do not allow one to unambiguously estimate the number of degrees of freedom, because such regularization imposes sparsity simultaneously over various spatial scales.

Nevertheless, our ensemble of six models allow us to qualitatively discuss how differences in model complexity can affect the resulting inference. In Sect.~\ref{ssec:results_lens} we compare the blindly submitted posterior distributions, and notice some differences in the inferred parameters uncertainties (in terms of the posterior standard deviation). Based solely on the reconstructed sources (see Fig.~\ref{fig:sim_data_comp_models}), the model that displays the least complex features is Sérsic+Shapelets. Oppositely, one of the models that capture the most complex features is Cluster+Exp+Lumweight. One may naively expect the Cluster+Exp+Lumweight model to be more affected by issues related to over-fitting and local minima, as it is defined on a locally fine grid and has more flexibility compared to the Sérsic+Shapelets. In case of over-fitting, the uncertainties on model parameters can be significantly under-estimated, partly due to discretization biases that artificially narrows the likelihood profile \citep[e.g.,][]{Nightingale2015,Etherington2022}. Interestingly, while Cluster+Exp+Lumweight indeed lead to overall small uncertainties on lens potential parameters (compared to other models, see Fig.~\ref{fig:sim_data_comp_lens_uncert}), we see that they do not significantly differ from the Sérsic+Shapelets model. Therefore, while we observe variations among the models regarding their posterior uncertainties (up to a factor of $\sim 1.6$), these variations are not solely driven by differences in model complexity. This result also highlights the difficulty in measuring the complexity of model simply based on the modeling assumptions and forms of regularization, or on the visual impression of the reconstructed source. However, this kind of analysis offers directions to explore for better understanding the origin of systematic biases in lens potential parameters.

\subsection{Model degeneracies between the lens and the source \label{ssec:mst_spt_discussion}}

As is presented in Sect.~\ref{ssec:lens_source_correl}, we observed correlations between the level of recovery (the relative error) of lens and source properties. In particular, we find a correlation between the lens power-law slope $\gamma$ and the source effective $r_{\rm eff}$, with correlation coefficient $r=0.9\pm0.4$, namely a positive correlation with a statistical significance of $2.3\sigma$. It is well-known that correlations between the source scale and the mass density profile of the lens can arise, which are often seen as a manifestation of the MSD \citep{Falco1985}. The MSD originates from the mass sheet transformation (MST), and its consequence is as follows: the addition or subtraction of a infinitely thin mass sheet from a power-law mass density profile changes, to first order, the slope of the density profile at the Einstein radius \citep[e.g.,][]{SchneiderSluse2013,Blum2020,Birrer2020}, while rescaling the source proportionally. A positive mass sheet locally increases the density slope, which in turns induces an increase in the source size through the MSD. This effect has also been empirically explored in \citet{Birrer2016} by using the explicit scale encoded in a source model based on shapelets. The positive correlation we find between the biases in $\gamma$ and $r_{\rm eff}$ may thus be the signature of the MSD: all the lens models we consider infer slightly too large density slopes and source effective radii.

There is a general trend of correlations between source light shape and lens mass shape that we observe in our results. For instance, we note correlations between the recovery of $q_{\rm s}$ and $q_{\rm m}$ or $\phi_{\rm ext}$, $r_{\rm eff}$, and anti-correlations with $\phi$ and the lens centroid. We also observe strong correlations between the error on $q_{\rm s}$ and that of the lens centroid. These correlations are more challenging to unambiguously interpret compared to those associated with the MST. We argue that they may be related to the source position transformation (SPT) outlined in \citet{SchneiderSluse2014} and further developed in \citet{Unruh2017}. The latter work effectively shows that transformation of the source profile can be compensated by changes in ellipticity and radial profile of the lens mass distribution. The biases that we observe in our models may be manifestations of this transformation and qualitatively match the expectation for a SPT. Because the SPT leads to an approximate degeneracy (unlike the MSD), it is expected to be broken with higher quality data. It may therefore be interesting to see how these correlations change with higher S/N data or even noise-free mock data.

More generally, approaches that are independent of a specific choice of mass model offer a complementary route to further study the interplay between lens and source structure. As reviewed in \citet[][and references therein]{Wagner2019}, instead of the direct modeling of imaging data pixels, a careful extraction of specific lensing observables can provide constraints onto local lens properties, without assuming global functional form of the lens potential. While such an approach is scale-independent by construction---it does not rely on a priori assumptions on the nature of the lensing object (isolated galaxy, cluster member, etc.)---it requires the ability to unambiguously extract lensing features like well-defined multiple images, their shape and orientation, or individual clumps within lensed arcs \citep[e.g.,][]{Wagner2018_cluster}. In galaxy-scale strong lenses such as the simulated one of Fig.~\ref{fig:simulated_data}, the observed images take the form of relatively smooth highly distorted arcs (i.e., rings), for which it is error-prone to attempt extracting local features of multiply imaged regions of the source without performing full lens modeling \citep{Galan2024corrfield}. Nevertheless, the local lensing formalism can be expanded to lensed arcs with globally smooth surface brightness, as developed in \citet{Birrer2021}. Such an approach is well-suited to make use of the multiple elongated arcs observed around massive galaxy clusters, as a way to locally correct the global mass model \citep[e.g.,][]{Yang2020}.

\subsection{Combining methods to mitigate systematic biases \label{ssec:post_combination}}

We quantify the scatter associated with the choice of modeling method in Table~\ref{tab:scatter_models}, which records the mean and scatter of relevant source light and lens mass quantities among all the models. Although dependent on the type of data considered in this work, these numbers provide an estimation of systematic errors that one can expect from modeling imaging data with different methods. The systematic errors we quote here can be interpreted as lower bounds to those of a real-case scenario, since we placed ourselves in an idealized setting -- no contamination by the lens light and perfect knowledge of noise properties and mass model family -- which removes a subset of the known sources of biases or degeneracies. Ideally, the addition of such complicating factors may broaden the posterior distributions from individual models, thus making them statistically compatible (underlying systematic biases would then be unnoticeable). However, more realistically, the scatter between models is likely to increase as a consequence of more complex models potentially subject to different sources of biases. Performing similar analyses as the one presented here to a wider variety of strong lensing data (different resolution, S/N, lensing configuration, nature of the source, etc.) will help understand the generalization of our results. For this purpose, the framework we have developed should help and encourage the multiplication of such analyses on both simulated and real data sets.

As we show in Fig.~\ref{fig:sim_data_comp_lens_full_corner} and Table~\ref{tab:scatter_models}, combining together the results from an ensemble of methods using uniform weights removes the observed biases. While individual methods display systematic errors in different parameters---and not necessarily always for the same parameters and in the same directions---it is reassuring that overall, we do not observe a significant residual offset after combination. Only for the mass density slope ($\gamma$) one can observe a global trend towards larger values. We quantify in Table~\ref{tab:scatter_models} the bias reduction between the biases from individual models and the one from the combined model. Among the eight mass model parameters, we find that an average bias reduction of 5.4, which is a substantial improvement on inferences from  individual models alone. This result is reassuring and shows that analyzing a given data set using independent modeling methods is an efficient way to mitigate systematic errors.

In this work, we conservatively assumed equal weights when combining individual methods. This is similar to the recent work of \citet{WongShajib2024}, where the authors combined measurements of $H_0$ obtained from by modeling the same lensed quasar with two different modeling codes, assuming equal weights. Nevertheless, we note that other approaches exist. The least conservative approach would be to multiply individual posteriors together, which is equivalent to assume that all methods are independent from each other and do not share systematic errors. The six modeling methods are only partially independent, as some of the models were performed by the same modelers using the same modeling code (see Appendix~\ref{app:sec:model_details}) albeit with different source reconstruction techniques. Additionally, we modeled a single imaging data realization of our simulated strong lens, which further introduces some degree of statistical dependence. The most Bayesian approach would be to combine individual models based on their Bayesian evidence. However, for reasons we detailed in Sect.~\ref{ssec:model_complexity}, objectively ranking models obtained from methods based on fundamentally different assumptions is still an unresolved issue. Nonetheless, better quantifying how different strong lens modeling methods perform on identical or similar datasets with respect to their complexity will be important to design better statistical combination procedures.

Beyond the accuracy improvement of a combined posterior distribution over the lens model parameters, the comparison of multiple models---and if possible, truly independent models---is extremely valuable for detecting unknown sources of systematics, updating our current modeling assumptions and techniques and testing additional models with different levels of complexity. Similarly, in source plane, comparing the morphology of different versions of the unlensed object (which is never directly observable) is necessarily valuable for straightening the confidence in a given feature, which in turn improves the robustness of its interpretation. Moreover, different models may predict different observables that may otherwise be overlooked, enabling the potential detection of anomalous lens systems \citep[such as missing images,][]{Ertl2024} or anticipate the detection of additional images unobserved in current observations \citep[e.g., similar to the geometric confirmation of multiple images in cluster-scale lenses,][]{Diego2023}.

\subsection{Extrapolating to time-delay applications \label{ssec:time_delay_app}}

\begin{figure}
    \centering
    \includegraphics[width=\linewidth]{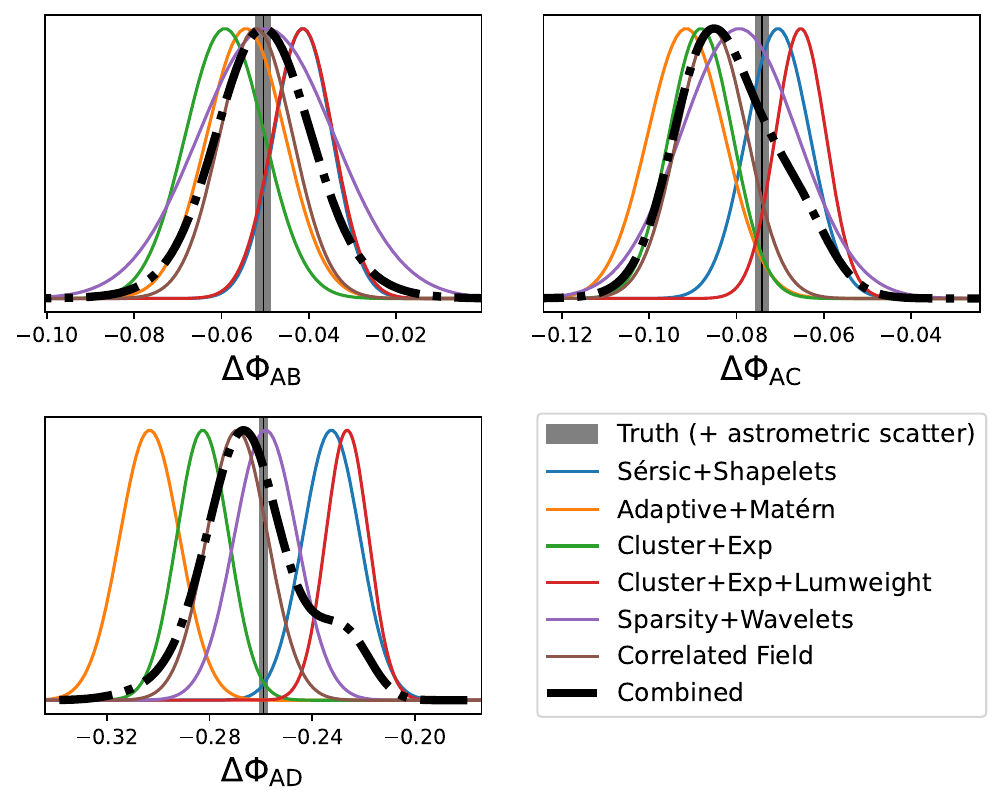}
    \caption{Fermat potential differences evaluated at image positions of a multiply point source located at the center of the source galaxy. Given that the modeled data (Fig.~\ref{fig:simulated_data}) does not contain the multiply imaged point source, this can be seen as an intermediate scenario between that of a lensed quasar (i.e., centered on its host) and a lensed SN (i.e., has faded away).}
    \label{fig:fermat_pot_diff}
\end{figure}

As lens modeling is a key ingredient in time-delay cosmography applications, one may be interested in the propagation of our results to the recovery of the Fermat potential difference (defined in Eq.~\ref{eq:fermat_pot_diff}). The Fermat potential difference between two images $i$ and $j$ is proportional to the time delay measured if the source is varying in time. Such a source can be a quasar centered on its host galaxy and orders of magnitude brighter than the lensed arcs, or supernova (SN) which can appear almost anywhere in its host and fades away after the explosion, leaving behind only the arcs as in our simulated data. 

We show in Fig.~\ref{fig:fermat_pot_diff} the Fermat potential differences for three hypothetical pairs of lensed images. While the simulated data we analyze in this work (Fig.~\ref{fig:simulated_data}) mimics the case of a lensed SN that faded away, we assume for simplicity that its location in source plane coincides with its host galaxy (although one could also proceed similarly for any other source position that leads to at least two lensed images). We label the lensed images ABCD, order them by ascending Fermat potential (i.e., $\Phi(\boldsymbol{\theta}_A)$ is the lowest) and consider the three independent pairs $ij\in\{{\rm AB, AC, AD}\}$. In addition to the posterior mass parameters uncertainties, we add in quadrature an additional uncertainty to mimic the limited astrometric precision \citep[for details, see][]{BirrerTreu2019}, assuming a conservative precision of 10 mas in image plane. The shaded gray area in each panels of Fig.~\ref{fig:fermat_pot_diff} around the true values isolates the typical contribution of the astrometric scatter term, which is in our case small compared to the uncertainties modeling the extended lensed source. As is expected from the constraints on mass profile parameters (Fig.~\ref{fig:sim_data_comp_lens_full_corner}), we observe a similar scatter around the true values among the models.

Interestingly, we do not observe a clear correlation between model biases on individual lens potential parameters and their resulting biases on Fermat potential differences. For example, the Sparsity+Wavelets model, which display slight biases on some source properties and lens potential parameters (e.g., the lens position), give Fermat potential posteriors well-centered on the expected values. While a robust generalization of these results require further work (e.g., on a larger sample of simulations), the general trend we previously found still remains: the combination of individual posteriors assuming equal weights results in posteriors that are free of biases. These combined distributions are shown with dash-dotted black lines in Fig.~\ref{fig:fermat_pot_diff}, and encapsulate well the Fermat potential difference values. This further gives motivation for time delay cosmography analyses to compare and combine lens models together, in particular those obtained using independent modeling methods \citep[see e.g.,][]{ShajibWong2022,Kelly2023}. In a follow-up paper, we shall explore the dependency of systematic lens modeling errors on the source plane position of the time-delay background object.

\subsection{Application to real data sets}

For the purpose of demonstration, we focus on simulated imaging data, but the framework and ideas we present is intended to be applied to real data. The COOLEST standard, which provides a common ground to express lens models from different origins, already supports real data and their corresponding models \citep[see e.g.,][]{Galan2024corrfield}. Moreover, extending the standard to other types of models (for both the lens and the source) is straightforward. One complication for applying the same framework presented here to real data arises from the human time and computation time required to apply multiple codes and methods to a given data set (see also Sect.~\ref{app:sec:model_details}).

Although COOLEST significantly reduces the burden to express a lens model in a format that can be readily compared to other models, the acquisition of combined posterior distributions over lens model parameters still requires that at least two distinct lens models are in hand. Here we model the same data using four software packages and six modeling methods, which is likely unrealistic in most real scenarios, especially given the large amount of lenses that are still to be modeled \citep[e.g., in the archival HST data, see][]{Garvin2022} and that will be discovered in the near future \citep[e.g.,][]{Collett2015}. Nevertheless, it is reasonable to assume that it is feasible to apply two or three methods to the same data (not necessarily within the same analysis), since the amount of lens modeling experts naturally grow over time and inference methods are becoming less and less time consuming. Deep learning methods, which we do not explore in the present work, offer promising avenues to accelerate the overall procedure either by proposing preliminary models to be refined with classical methods, or providing additional posterior distributions for final combination, at a negligible cost (ignoring the training phases).

When modeling real data, the chosen mass model parametrization is only an approximation of the true mass distribution. Extending the present analysis to account for differences between the truth and the model is obviously conceivable. However, it is important to note that such an extension inherently requires key assumptions in order to properly define ``how different'' is the truth compared to the model. A concrete example is the work of \citet[][Sect.~3.2]{Gomer2022} who simulated a population of SLACS-like strong lenses following a composite (baryons + dark matter) mass model, for which they had to assume a series of additional assumptions (ellipticity, halo scale radius, etc.), which they subsequently modeled with a power-law profile radially modified with a mass sheet parameter. \citet[][Sect.~2.4]{Cao2022} simulated strong lenses with mass distribution based on stellar kinematics models and performed modeling using single power-law profiles with shear. \citet{VanDeVyvere2022} quantified the biases caused by a lack of freedom along the azimuthal direction, when the true mass distribution contains multipoles or ellipticity gradients. The Rung 3 of the TDLMC, although focused on the recovery of $H_0$ instead of mass model parameters, has been a blind experiment where independent teams modeled lenses extracted from cosmological simulations \citep{Ding2021}. On real and simulated systems, the Time-Delay Cosmography collaboration \citep{Millon2020} systematically uses both power-law and composite mass model families and marginalizes over the resulting posteriors, which are further constrained along the radial direction with stellar kinematics information \citep[e.g.,][]{ShajibWong2022}. Given the diversity of deviations to the commonly assumed sheared power-law elliptical mass profile in real galaxies---such as truncations \citep[e.g.,][in dense environments]{Limousin2007}, cores \citep[e.g.,][]{Collett2017} or multipoles \citep[e.g.,][]{Stacey2024}---the generalization of the above-cited results to all strong lens systems is still unclear. It is possible that for a given real system, a set of dedicated simulations based on a set of likely mass models, followed by lens modeling as for the real data, may be the best way to properly quantify systematic errors caused by the choices of mass model.

\section{Conclusion \label{sec:conclusion}}

We have conducted a fully blind modeling experiment on strong lensing data simulated with a dedicated software package by one author, while four other authors used four independent software packages to model it. In total, six modeling methods---which differ in their source reconstruction techniques and inference pipelines---have been applied on that same data. We have made a series of simplifying assumptions to keep this novel kind of analysis tractable, in particular regarding the lens light, the mass model parametrization and the noise properties. In contrast, the true PSF, lens parameters and source morphology were hidden and the optimization and inference strategies left free to the modelers.  We used the lensing standard COOLEST to overcome the challenges that arise when comparing results obtained with different modeling codes. The resulting image and source plane models, as well as model residuals are given in Fig.~\ref{fig:sim_data_comp_models}, and constraints on lens potential parameters are shown in Fig.~\ref{fig:sim_data_comp_lens_full_corner}. Below we summarize our main results:

\begin{itemize}
\setlength\itemsep{0.8em}
    
    \item While no modeling method resulted in strong statistical biases systematically for all lens and source properties, we observed a measurable scatter among the models. Strongest biases arise for the lens centroid, while the mass density slope at Einstein radius is only mildly biased with a small inter-model scatter. We also observe differences in the dynamic range of the reconstructed source intensities.

    \item Combining results from different modeling techniques enables to mitigate systematic uncertainties arising for individual models. For the particular data we consider, the systematic error on lens potential parameters is reduced by a factor 5 on average. The reason is that models tend to scatter around the true parameters values but stay statistically compatible, such that the combined posterior distributions effectively broaden and include the true values. This results also holds regarding the Fermat potential differences between hypothetical lensed images of a point source component, which is relevant for time delay cosmography applications. While the amount of bias reduction is evidently data-dependent, we argue that model combination is generally beneficial to remove some biases from strong lensing analyses.

    \item Towards the goal of better understanding the origin of model biases, we used our ensemble of models to investigate possible correlations between lens and source properties. We observed correlations between errors on lens potential parameters (e.g., the mass density slope) and on the morphology of the source (e.g., the effective radius or axis ratio). We argue that such correlations can be manifestations of the well-known mass-sheet transformation (MST), but also more generally of the source-position transformation (SPT). Better handling these degeneracies in current and future modeling methods will be key to further minimize model biases.

    \item We investigated how certain model assumptions affect the recovery of lens potential parameters. In particular, we explore (1) how the cuspy nature of the lensed galaxy can lead to systematic errors if the source model is not flexible enough to capture large intensity variations, and (2) how the accuracy of the PSF (the true PSF being unavailable to the modelers) plays a role even for extended source modeling. We find that both an accurately sampled PSF and a source model with large dynamic range (e.g., using a luminosity-weighted source prior) are warranted to reconstruct cuspy lensed sources while minimizing systematic errors on lens potential parameters.
    
\end{itemize}

Over the past years, numerous lens modeling methods have been proposed and implemented in different software packages. Here, we selected a subset of those with the goal of using them together, instead of only opposing them. Typically, we refrain from explicitly ranking the modeling methods, which would only be meaningful over an extremely large sample of strong lenses with different data quality and modeling assumptions to ensure proper generalization. In a real-case scenario, we do not have access to the truth; therefore, combining the results from different methods is a pragmatic and efficient way to detect and mitigate systematic errors. As is shown, for example, in \citet{Suyu2006} by testing three types of source regularizations, there exists an inherent dependence on the data properties, the lens configuration and the (unobservable) intrinsic source morphology, such that it is unlikely that a single source reconstruction method gives unbiased results in all cases. Our work strengthens this idea and goes further by combining a large collection of models, while investigating their specific effects on inferred parameters. Moreover, we purposefully allowed some level of freedom for the modelers (e.g., masks, PSF, posterior marginalization, etc.), such that our work also illustrates the role of specific modelers' choices. These choices play a role in the observed scatter between models, and can be marginalized over by combining the methods together.

As stated in the introduction, we have not used lens modeling methods based on deep learning, as those would require many additional assumptions, in particular regarding their training phase. Nevertheless, the comparison framework presented here is very general and does not depend on the nature (classical, deep learning, etc.) of the underlying methods. Therefore, we encourage future studies to complement and combine classical methods with those based on deep learning, as the latter have clear advantages such as fast computation time after training and a large flexibility through different network architectures. Examples of using neural networks to complement classical techniques have been proposed in \citet{Maresca2021} and \citet{Pearson2021_comb_nn}. Our publicly released simulated data and lens modeling products can be directly used to test and improve such deep learning (or any other) approaches.

The framework and ideas presented here is designed to be applied on real data sets and expanded beyond our initial simplifying assumptions. In particular, the role of the lens surface brightness model should be investigated further \citep[see e.g.,][in the context of subhalo detection]{Nightingale2024}. Similarly, the standard assumptions of uncorrelated Gaussian noise used in lens modeling analyses should be re-assessed \citep[e.g., recent JWST imaging data show strongly correlated noise patterns, see][]{Rigby2023} to ensure that analyses of the many future observations of strong lenses remain fully accurate.

\begin{acknowledgements}
The authors thank the anonymous referee for reviewing the original manuscript and providing useful comments. This work originated in the Lensing Odyssey 2021 workshop\footnote{\url{https://lensingodyssey.com}}, and so we would like to acknowledge the organizers and attendees for the fruitful discussions. 
AG acknowledges the Swiss National Science Foundation (SNSF) for supporting this work. This work was also supported by the European Research Council (ERC) under the European Union’s Horizon 2020 research and innovation programme (COSMICLENS: grant agreement No 787886). GV and QM were both supported by the generosity of Eric and Wendy Schmidt by recommendation of the Schmidt Futures program. QM gratefully acknowledges a grant of computer time from ACCESS allocation TG-AST130007.
This research has made use of \textsc{SciPy} \citep{Virtanen2020scipy}, \textsc{NumPy} \citep{Oliphant2006numpy,VanDerWalt2011numpy}, \textsc{Matplotlib} \citep{Hunter2007matplotlib}, \textsc{Astropy} \citep{astropy2013,astropy2018}, \textsc{Jupyter} \citep{Kluyver2016jupyter} and \textsc{GetDist} \citep{Lewis2019getdist}.
\end{acknowledgements}


\typeout{}  
\bibliographystyle{aa}
\bibliography{biblio}


\begin{appendix}

\section{Realistic source surface brightness in mock data \label{app:sec:previous_mock}}

\begin{figure}
    \centering
    \includegraphics[width=\linewidth]{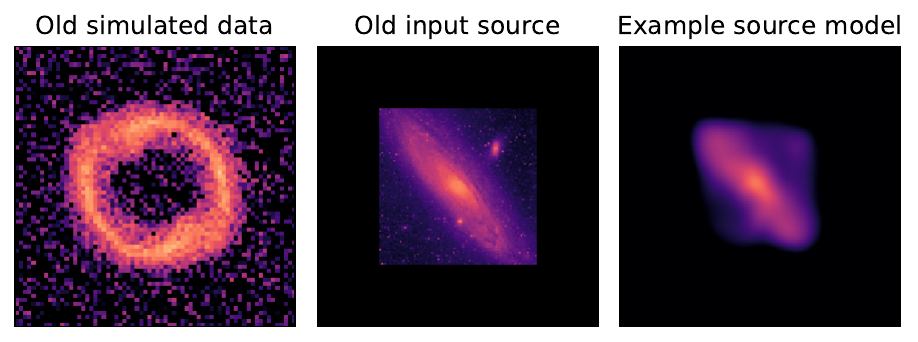}
    \caption{Old simulated lensing data based on another source galaxy. For that mock the source image background was insufficiently subtracted, which left faint imprints in the data after being lensed. While unnoticeable to the eye in the data (left panel), the square boundaries of the source are clearly noticeable in the input source (middle panel) and even on the reconstructed source (right panel, using a correlated field model as an example). These artificial boundaries led to biases in lens potential parameters and source size. Note that a logarithmic scale is used in all panels.}
    \label{app:fig:old_simulated_data}
\end{figure}

The mock data shown in Fig.~\ref{fig:simulated_data} is not the first mock we built for this work. Initially, an other prescription for the background source was used, which caused large biases on many model parameters, which was noticed only after unblinding. Therefore, it was decided to repeat the entire procedure---namely, the data simulation and blind submissions of models from independent modelers---which lead to the mock used throughout this work. As it can be beneficial to some readers, we give more details about the old mock data below and why it caused large parameters biases.

We show in the left panel of Fig.~\ref{app:fig:old_simulated_data} the old simulated data. For the source, shown in the middle panel, we used a B filter image of M31 available through the ESO Online Digitized Sky Survey. This image was selected because a local galaxy has high-detail resolved structure with negligible PSF spikes, which could introduce nonphysical features if lensed. After ray-tracing and the addition of noise, the data visually resembles to genuine lensing data, despite the input source boundaries being visible at some locations. Therefore, any potential issue related to these boundaries could not be detected before modeling the data.

After the blind modeling stage, it was noticed by the modelers that their reconstructed source models were displaying strong boxy features surrounding the ellipsoidal shape of the source galaxy. Among the six models, the Sparsity+Wavelets model was strongly affected by the square and boxy nature of the source, for which an extremely low resolution grid was systematically preferred as it represented better these sharp boundaries. This artificial bias towards a low resolution grid (i.e., few source pixels) was in turn strongly biasing the lens potential parameters, in particular the mass density slope due to its degeneracy with the source size (through this MST). The right panel of Fig.~\ref{app:fig:old_simulated_data} shows, as an example, the Correlated Field model of that source, which also displays a boxy shape. After comparing the different models together, the mass density slope was found to be biased in almost all models. After extensive checks, the origin of these biases was attributed to the sharp boundaries and large background flux of the input source. The source used in the new simulated data (shown in Fig.~\ref{fig:simulated_data}) was carefully processed, which solved these issues.

\section{Technical modeling details \label{app:sec:model_details}}

\begin{table*}[h!]
    \caption{Computation times for the six modeling methods, including best-fit optimization and posterior distribution sampling.\label{app:tab:runtimes}}
    \renewcommand{\arraystretch}{1.3}
    \centering
    \small
    \begin{tabular}{llllll}
         Modeling method label & Software package & Involved steps & Longest instance & Total & Hardware configuration \\
         \hline\hline
         Sérsic+Shapelets & \lenstronomy & PSO + MCMC & 17 hours & 38 hours & Intel Core i7 CPU (2 cores) \\ 
         Adaptive+Matérn & \vkl & Nested sampling & 4 days & 4.2 days & \tiny{Intel Xeon Platinum 8260 CPU (96 cores)} \\
         Cluster+Exp & \qlens & Nested sampling & 1.5 hours & 1.5 hours & \tiny{Intel Xeon Platinum 8160 CPU (192 cores)} \\
         Cluster+Exp+Lumweight & \qlens & Nested sampling & 2 hours & 2 hours & \tiny{Intel Xeon Platinum 8160 CPU (192 cores)} \\
         Sparsity+Wavelets & \herculens & Grad. descent + SVI & 6 minutes & 3.4 hours & NVIDIA A10 GPU \\
         Correlated Field & \herculens & Grad. descent + geoVI & 45 minutes & 4.5 hours & NVIDIA A10 GPU \\
    \end{tabular}
    \tablefoot{
    Some of the final models were obtained by combining several model instances, so we quote the runtime for a single instance (the most expensive) in addition to the total runtime (assuming single instances were not performed in parallel). For additional details and software package references, see Sect.~\ref{sec:modeling_methods} and Appendix~\ref{app:sec:model_details}.
    }
\end{table*}

This section gives technical details on the specific choices made by modelers to analyze the simulated data. We complement it with Table~\ref{app:tab:runtimes} which gives the approximate computation time necessary to obtain the various models used in this work.

\subsection{S{\'e}rsic+Shapelets \label{app:ssec:sersic_shapelets}}

For the analytical model using \lenstronomy(Sect. ~\ref{ssec:general_lenstronomy}), a 4.5$\arcsec$-radius circular mask around the lens is used. The source light is composed of a single elliptical S{\'e}rsic profile. Shapelets components are added sequentially to that profile. The significance of those components to the model is evaluated by calculating the  Bayesian Information Criterion (BIC),  which balances the likelihood with the number of parameters following:
\begin{equation}
\label{eq:def_bic}
    {\rm BIC} = n_{\rm par}  \ln({n_{\rm data}}) - 2  \ln{L(\tilde{\boldsymbol{\eta}})} \ ,
\end{equation}
where $n_{\rm par}$ is the number of parameter, $n_{\rm data}$ is the number of data points (i.e., data pixels used as constraints), and $L(\tilde{\boldsymbol{\eta}})$ is the loss function evaluated at the best fit position (Eq.~\ref{eq:loss_func}). For this specific mock, the BIC favors a maximum order of shapelets $n_{\rm max} = 5$ (Table \ref{tab:shapelet_order}).

\begin{table}[]
    \caption{Comparison of best-fit log-likelihood ($\ln{L(\tilde{\boldsymbol{\eta}})}$) and corresponding BIC values (Eqs.~\ref{eq:loss_func} and \ref{eq:def_bic}) considering different maximum order of the shapelets basis set $n_{\rm max}$ for the Sérsic+Shapelets modeling method (and correspinding number of parameters $n_{\rm par}$). The number of data pixels is 6376 used to constrain the model (within a circular mask of $4\farcs5$ radius).}
    \label{tab:shapelet_order}
    \renewcommand{\arraystretch}{1.3}
    \centering
    \begin{tabular}{cccc}
        $n_{\rm max}$ & $n_{\rm par}$ & $\ln{L(\tilde{\boldsymbol{\eta}})}$ & BIC \\
        \hline\hline
         3& 28 & -3581.4 & 7408.0  \\
         4& 33 & -3562.8 & 7414.7\\
         5& 39 & -3485.1 & 7311.9\\
         6& 46 & -3502.2 & 7407.1\\
         7& 54 & -3434.2 & 7341.4
    \end{tabular}
\end{table}

The point spread function (PSF) is treated in \lenstronomy as follows. The surface brightness of the lensing system, which is simulated at each iteration, can be sampled on a grid with higher resolution than the observed image before being averaged to the data resolution. If the modeled surface brightness is supersampled, the user can perform the PSF convolution on the finer grid. In that case, a supersampled PSF is calculated by interpolation, and an iterative process allowing for perturbation of individual PSF pixels, ensures that the downsampling of the supersampled PSF recovers the input PSF. We elected a supersampling factor of 5 and performed the PSF convolution on the supersampled grid.

\subsection{Adaptive+Matérn}
\label{app:ssec:ag_vor_mat}
We begin with the broadest possible range of the lens potential parameters of the model (see Eqs. \ref{eq:def_pemd} and \ref{eq:def_extshear}), except for the slope, which we fix to isothermal, and the Einstein radius, whose range is fixed by rough estimation of the radius of the Einstein ring from the data.
The former serves simply to speed up the calculation, while the latter is required to avoid over- or under-focused, non-physical solutions.
The main choices at this first-pass stage are the type of regularization and the resolution of the adaptive grid.
For the regularization we choose curvature, which has only one free parameter, the regularization strength $\lambda$, and a fixed covariance matrix $C_s$ (see Eq. \ref{eq:quadratic_reg_general}).
The adaptive grid is created simply by using the deflected positions of every third pixel of the data image in the x and y directions.
We create a custom mask that we use in all of our subsequent models, and we do not perform any supersampling of the PSF.
This setup is very fast to run, taking only a few minutes on a standard multi-core laptop.

After this first pass, we switch to using the Matérn kernel regularization given in Eq. \ref{eq:mattern}, which has 3 free parameters, and increase the resolution of our adaptive grid by using every second pixel in both directions in the data image to construct it.
Based on our first crude model, we restrict each lens potential parameter to about 20 per cent of its full range, centered roughly on the bulk of the posterior probability.
This model takes about an hour to run on a standard multi-core laptop.

Finally, we further restrict the parameters to 10 per cent of their full range and run a final model with the same regularization but even higher resolution; that is, shooting back every pixel in the data image to create the adaptive grid.
We combine the last two models, which differ only in the resolution of the reconstructed source (although the first served to initialize the second in order to save on computations), by merging their posterior samples in such a way that the probability mass of each one is the same independently of its actual size.

\subsection{Cluster+Exp and Cluster+Exp+Lumweight \label{app:ssec:ag_vor_clust}}

In \qlens, the image plane is supersampled by splitting each image pixel into $3\times 3$ subpixels; each subpixel is ray-traced to the source plane, and the source grid is generated using a $k$-means clustering algorithm exactly as is done in \citet{Nightingale2018}, but with each ray-traced point receiving equal weight.  We choose the number of source pixels to be equal to half the number of image pixels within the mask. The surface brightness of each ray-traced subpixel is determined by interpolating in the three source pixels whose Delaunay triangle the ray-traced point is in (or closest to); the surface brightness for all the subpixels within a given image pixel are then averaged to obtain the surface brightness for the image pixel. The pixel surface brightness values obtained this way are then convolved with the pixel-level PSF (note that although we are supersampling the image plane, we do not supersample the PSF here; the effect of PSF supersampling will be explored in Section \ref{sec:supersampling}). The regularization is performed using an exponential kernel (equivalent to a Matérn kernel with $\nu = 1/2$).

To encourage convergence, we make use of two additional priors on the source: first, there is a prior that discourages producing lensed images outside the mask. We accomplish this by temporarily unmasking after the source pixels are solved for, generating the lensed images without a mask, and imposing a steep penalty if surface brightness is found outside the mask whose value is greater than 0.2 times the maximum surface brightness of the images. Second, we place a prior on the number of lensed images produced. This is accomplished by creating a Cartesian grid in the source plane and finding the overlap area of all the ray-traced image pixels for each Cartesian grid cell; by dividing the total overlap area by the area of each grid cell, we obtain the number of images produced by that cell. We can then take the average number of images over all the cells. We impose a steep penalty if the average number of images if less than 1.5. This discourages solutions that are not multiply imaged, where the source looks identical to the observed configuration of lensed images. With these priors in place, we can obtain a good solution with a single nested sampling run, provided the parameter priors are broad enough.

The Cluster+Exp+Lumweight model uses all of the methods described above, but in addition it uses a luminosity-weighted regularization, as is described in Sect.~\ref{ssec:general_semlinear_inv}. Thus we include the additional parameter $\rho$ (Eq.~\ref{eq:weightfunc}), which controls the steepness of the luminosity weighting, as an additional nonlinear parameter to be varied.

\subsection{Sparsity+Wavelets \label{app:ssec:ms_wavelets}}

Before modeling the source on a regular grid with multi-scale regularization, we start with an approximate lens mass model obtained by modeling the source with a single Sérsic profile. Since at this stage of the modeling process the mass model may be rather inaccurate, using spatially varying regularization weights (${\rm \bf W}_{\rm ms}$ in Eq.~\ref{eq:wavelet_regul}) could bias the source reconstruction. Therefore we approximate the weights by their median value within each wavelet decomposition scale (i.e., we use spatially uniform weights within each frequency range). We set the global regularization strength $\lambda_{\rm ms}=3\sigma$ for the first wavelet scale (highest frequency features), and $\lambda_{\rm ms}=1\sigma$ for the remaining scales. We choose a lower threshold for low frequency features as advocated in various works relying on similar multi-scale regularization strategies \citep[e.g.,][]{Lanusse2016,Peel2017,Galan2022herculens}, since high frequencies are more impacted by the presence of noise in the data.

We obtain a first approximate model of the pixelated source by jointly optimizing all parameters except for the fixed lens center, and we impose a strong isothermal prior (i.e., $\gamma\sim\mathcal{N}(2, 10^{-3})$) on the mass density slope to avoid introducing degeneracies early in the modeling sequence. We use the gradient descent optimizer \textsc{AdaBelief} \citep{Zhuang2020_AdaBelief} implemented in the \textsc{Optax} \citep{optax2020github} library to obtain best-fit parameters. We then re-optimize model parameters by re-computing regularization weights and releasing priors on the lens center and density slope.

The last step is to estimate the posterior distribution of lens mass parameters while further refining the source model. At this stage, the lens model is very close to the best-fit model so we do not rely anymore on the approximation of uniform regularization weights per wavelet scale, and properly propagate the noise to source plane. These more accurate regularization weights significantly help eliminating remaining artifacts located on the outskirts of the source galaxy, when the data is the least constraining. The resulting $\chi^2$ being below unity, we further boost the regularization strength of high-frequencies (for this particular data, by $5\sigma$) in order to obtain a $\chi^2$ of the order of unity. We note that the precise value of this boost does not significantly impact the final posterior distribution as we marginalize over many model variations. In particular, we vary the number of source pixels from $80\times80$ to $160\times160$ pixels with steps of 5. We also run the same ensemble of models by globally increasing the regularization strength by $1\sigma$. In total, we consider 34 model variations.

The joint posterior distribution for lens mass parameters is estimated using stochastic variational inference \citep[SVI, see review by][]{Blei2016_review_svi}, which directly makes use of known gradient of the loss function. SVI is also less computationally expensive than other sampling methods such as Markov Chain Monte Carlo or Hamiltonian Monte Carlo, which allows us to run a larger number of model variations. Since in this work, we are mainly interested in the joint posterior distribution of the lens mass parameters which are not expected to exhibit strongly non-Gaussian correlations\footnote{This assumption is validated with the correlated field model by using more flexible SVI methods (see Sect.~\ref{app:ssec:correl_field}).}, we find that a multi-variate Gaussian distribution is a sufficient surrogate posterior model. However, we acknowledge that SVI can underestimate uncertainties \citep[as in][]{Gu2022}, a limitation that we address by marginalizing over the 34 model variations assuming equal weights. In addition to the full posterior distributions and first-order posterior statistics, we save in \coolest format a point-estimate model, that corresponds to the mean model as obtained from SVI, with $120^2$ pixels and fiducial regularization strength.

\subsection{Correlated Field \label{app:ssec:correl_field}}

Similar to the strategy used with the multi-scale source model, we first model the imaging data with a single Sérsic source profile, which we then replace with the correlated field model defined in Eq.~\ref{eq:src_correl_field}. For our baseline model, we set the shape of the Gaussian excitation field to $90^2=8'100$ pixels. The field power spectrum is modeled as power law parametrized with an amplitude and a slope. These two parameters are themselves sampled from a log-normal distribution, described in turn by a mean and scale parameters. The additive offset in real space is modeled by a single scalar (initialized to the mean flux values from the Sérsic model), whereas variations in this offset are sampled from a log-normal distribution with additional mean and scale parameters. Lens potential model parameters (PEMD and external shear) are sampled from Gaussian distribution, for which we check that the prior widths are large enough. We optimize the full model---lens mass parameters and source field parameters---using the set of minimizers and samplers implemented in \nifty. More specifically, we converge to the maximum a posteriori solution and estimate the joint posterior distribution using metric-Gaussian variational inference \citep[MGVI,][]{Knollmuller2019_mgvi}.

We ran several variations of the above fiducial models. In particular, we increased the field resolution to $120^2=14'400$ pixels, alter the sampler random seeds, initialized the model with a worst lens model (obtained from a different Sérsic source model), and used geometric variational inference \citep[geoVI,][]{Frank2021_geovi} instead of MGVI. All these model variations result in almost identical posterior distributions for lens mass parameters and consistent source models. Nevertheless, we conservatively marginalize over these models with equal weights. For the point-estimate parameters, we set those to the mean values of the VI samples of the model with the most resolved source (i.e., $120^2$ pixels), although the fiducial model is virtually indistinguishable.

\section{Correlations between lens and source properties \label{app:sec:lens_source_correl}}

\begin{figure*}
    \centering
    \includegraphics[width=\linewidth]{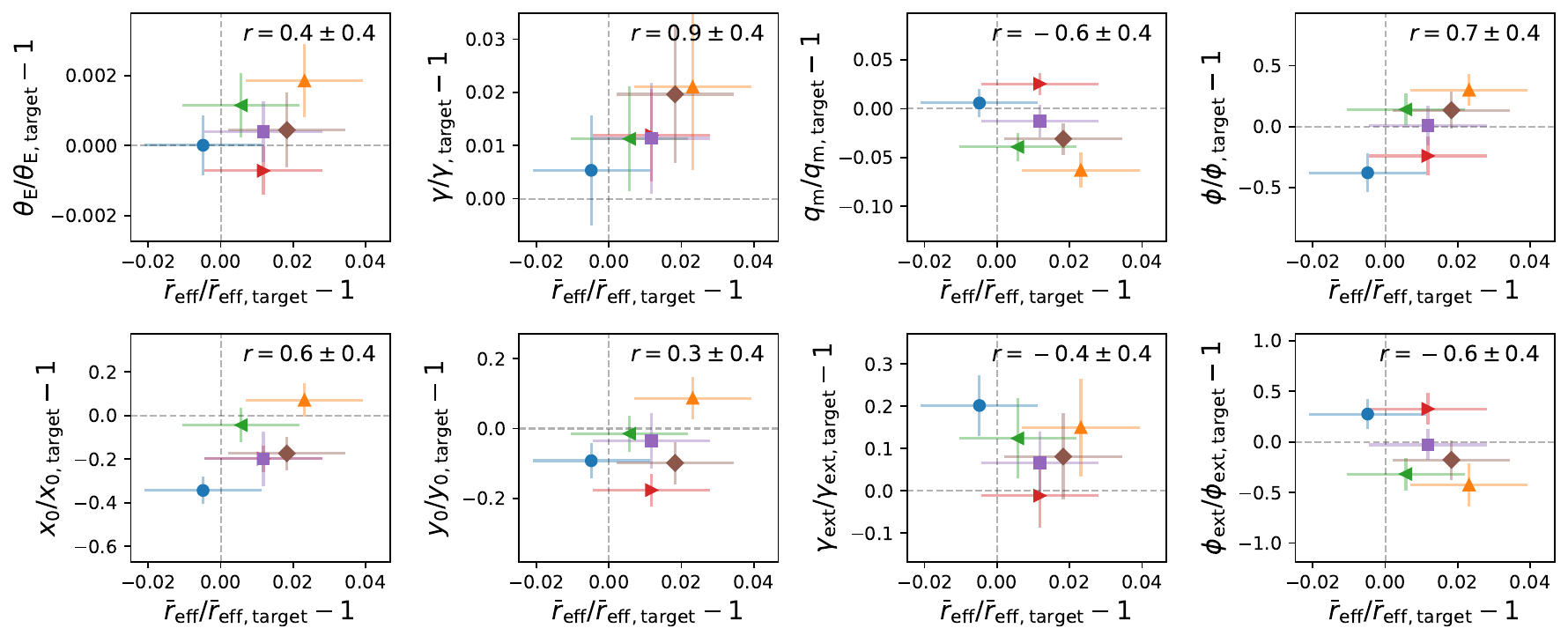}
    \caption{Relative error with respect to the true values. $x$ axis: source effective radius, $y$ axis: mass model parameters. The legend markers and colors are the same as in Fig.~\ref{fig:sim_data_comp_src_props}. In the top right part of each panel, the biweight mid-correlation $r$ is indicated (0 means no correlation).}
    \label{app:fig:correl_r_eff}
\end{figure*}
\begin{figure*}
    \centering
    \includegraphics[width=\linewidth]{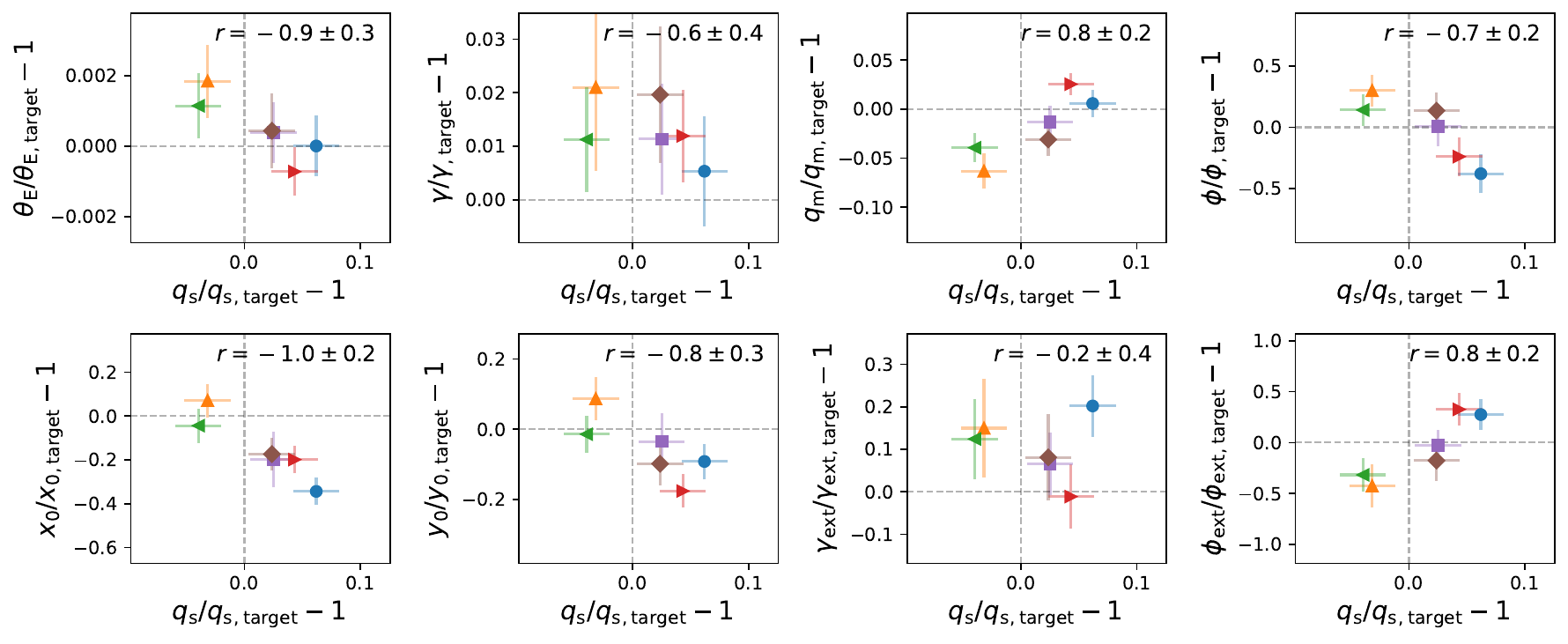}
    \caption{Relative error with respect to the true values. $x$ axis: source axis ratio (estimated from central moments), $y$ axis: mass model parameters. The legend markers and colors are the same as in Fig.~\ref{fig:sim_data_comp_src_props}. In the top right part of each panel, the biweight mid-correlation $r$ is indicated (0 means no correlation).}
    \label{app:fig:correl_axis_ratio}
\end{figure*}
\begin{figure*}
    \centering
    \includegraphics[width=\linewidth]{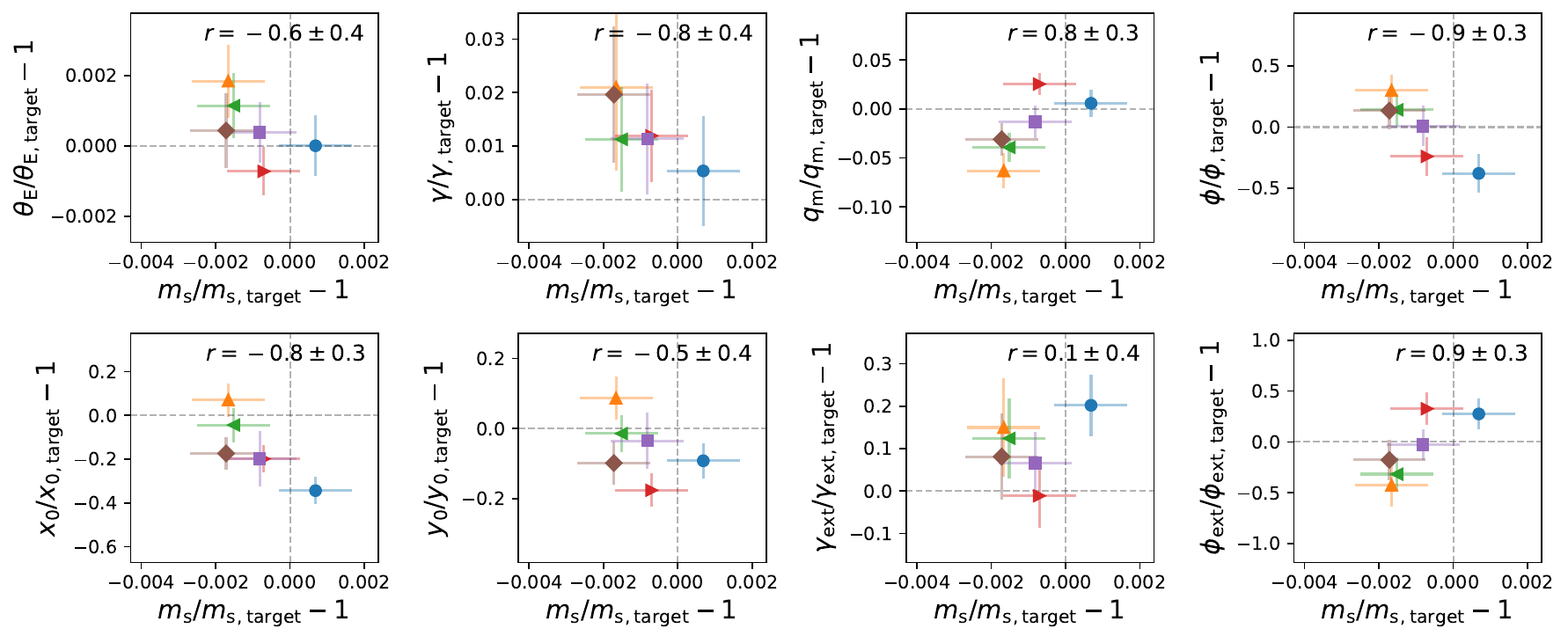}
    \caption{Relative error with respect to the true values. $x$ axis: source total magnitude, $y$ axis: mass model parameters. The legend markers and colors are the same as in Fig.~\ref{fig:sim_data_comp_src_props}. In the top right part of each panel, the biweight mid-correlation $r$ is indicated (0 means no correlation).}
    \label{app:fig:correl_mag}
\end{figure*}

Given the six independent models we gather in this work, we can investigate of the errors on the lens and source properties of interest correlate with each other. In other, we are interested in signs of degeneracies between the lens and source properties, that can be revealed over the ensemble of models. We show in Figs.~\ref{app:fig:correl_r_eff}, \ref{app:fig:correl_axis_ratio} and \ref{app:fig:correl_mag} a series of plots that correlate all lens potential parameters with the three main source properties we investigate ($r_{\rm eff}$, $q_{\rm s}$ and $m_{\rm s}$ respectively). These results are presented and discussed in the main text, in particular in Sect.~\ref{ssec:lens_source_correl} and Sect.~\ref{ssec:mst_spt_discussion}.

\end{appendix}

\typeout{get arXiv to do 4 passes: Label(s) may have changed. Rerun}

\end{document}